\documentclass[12 pt]{article}
\usepackage{putex}
\usepackage{graphicx}
\usepackage{xspace}

\newcommand{\AdS}{\ensuremath{\mbox{AdS}_5}\xspace}

\begin{document}

\preprint{PUPT-2269 \\ LMU-ASC 28/08}

\institution{PU}{Joseph Henry Laboratories, Princeton University, Princeton, NJ 08544, USA}
\institution{MaxPlanck}{Ludwig-Maximilians-Universit\"at, Department f\"ur Physik, Theresienstrasse 37, \cr 80333 M\"unchen, Germany}

\title{Entropy production in collisions of gravitational shock waves and of heavy ions}

\authors{Steven S. Gubser,\worksat{\PU,}\footnote{e-mail: {\tt ssgubser@Princeton.EDU}}
Silviu S. Pufu,\worksat{\PU,}\footnote{e-mail: {\tt spufu@Princeton.EDU}} and
Amos Yarom\worksat{\MaxPlanck,}\footnote{e-mail: {\tt yarom@theorie.physik.uni-muenchen.de}}}

\abstract{We calculate the area of a marginally trapped surface formed by a head-on collision of gravitational shock waves in AdS${}_D$.  We use this to obtain a lower bound on the entropy produced after the collision.  A comparison to entropy production in heavy ion collisions is included.  We also discuss an $O(D-2)$ remnant of conformal symmetry which is present in a class of gravitational shock wave collisions in AdS${}_D$ and which might be approximately realized (with $D=5$) in central heavy-ion collisions.}

\date{May 2008}

\maketitle
\tableofcontents

\section{Introduction and summary}
\label{INTRODUCTION}

Relativistic heavy ion collisions produce a lot of entropy.  Consider for example gold ions colliding with $\sqrt{s_{NN}} = 200\,{\rm GeV}$.  A head-on collision (usually described as ``central'') produces about $5000$ charged tracks: see for example \cite{Back:2004je}.  The total entropy may be roughly estimated as
 \eqn{Srough}{
  S \approx 7.5 N_{\rm charged} = 38000 \,.
 }
In section~\ref{PHENOMENOLOGY}, we explain where the factor of $7.5$ comes from.

The main aim of this paper is to inquire how well one can understand total entropy production in a heavy ion collision in terms of a dual black hole description. Ideally, we would like to construct colliding nuclei in a holographic dual to QCD\@.  When the duals of the nuclei collide in the bulk, a black hole should form, signifying the formation of a quark-gluon-plasma.  While a holographic dual to QCD is unavailable, it was suggested early on \cite{Nastase:2004pc,Nastase:2005rp} that an analogy should exist between colliding heavy ions and colliding gravitational shock waves in anti-de Sitter space.  Subsequent related work on collisions in \AdS includes \cite{Shuryak:2005ia,Amsel:2007cw,Grumiller:2008va}.  In the next few paragraphs, we will summarize an entropy estimate based on colliding gravitational shock waves which gives a result surprisingly close to \eno{Srough}.

The line element for two identical head-on shock waves propagating toward one another in \AdS is
 \eqn{AdSShocks}{
  ds^2 = {L^2 \over z^2} \left[ -du dv + (dx^1)^2 + (dx^2)^2 + dz^2
   \right] +
    {L \over z} \Phi(x^1,x^2,z) \left[ \delta(u) du^2 +
      \delta(v) dv^2 \right] \,,
 }
where we have introduced the light-cone coordinates
 \eqn{uvDef}{
  u = t-x^3 \qquad v = t+x^3 \,,
 }
and have assumed that $u<0$ or $v<0$.
A simple shock-wave geometry in anti-de Sitter space can be obtained by boosting a black hole solution. As we will explain in section~\ref{SHOCKS},
for such a shock wave, the function $\Phi(x^1,x^2,z)$ in \eno{AdSShocks} is given by
 \eqn{PhiDef}{
  \Phi(x^1, x^2, z) = {2 G_5 E \over L} {1 + 8 q (1+q) - 4 \sqrt{q (1+q)} (1 + 2 q) \over \sqrt{q (1+q)}}
 }
where
 \eqn{qDef}{
  q \equiv {(x^1)^2 + (x^2)^2 + (z-L)^2\over 4 z L}
 }
and $E$ is the total energy of the shock wave.
In the rest of this introductory discussion we focus on shock waves of the form \eqref{PhiDef}.  Extensions to more general shock waves in various dimensions can be found in sections \ref{DIMENSIONS} and \ref{OTHER}.

The metric \eno{AdSShocks} has singularities at $u=q=0$ and $v=q=0$ where Einstein's equations apply only in a distributional sense.  These singularities merely signal the presence of pointlike massless particles of energy $E$, remnants of the boosted black hole.  These singularities could be smoothed out by replacing each massless particle by a continuous cloud of massless particles with the same total energy. In \cite{Giddings:2004xy}, point-like sources for shocks propagating in a flat-space background were replaced by wave-packets.  The geometry \eno{AdSShocks} describes a head-on collision because the massless particles are located at the same position in the transverse space parameterized by $x^1$, $x^2$, and $z$.

The reason we must assume $u<0$ or $v<0$ in \eno{AdSShocks} is that the two shocks collide at $u=v=0$, and in the future light-volume of that event, little is known about the geometry (see however \cite{Grumiller:2008va}.) Assuming a black hole is formed after the collision, there is a standard method \cite{Penrose,DEath:1992hb,DEath:1992hd,DEath:1992qu} for computing a lower bound on the entropy $S$ of the black hole:
 \eqn{PenroseBound}{
  S \geq S_{\rm trapped} \equiv {A_{\rm trapped} \over 4 G_5} \,,
 }
where $A_{\rm trapped}$ is the area of the trapped surface: that is, a surface whose null normals all propagate inward.  The inequality \eno{PenroseBound} is based on the expectation that trapped surfaces must lie behind an event horizon.  To our knowledge, \eno{PenroseBound} has not been rigorously demonstrated in anti-de Sitter space.  It is related to singularity theorems \cite{Hawking:1969sw}, cosmic censorship (for a review see \cite{Penrose:1999vj}) and the area theorem, which is usually proven on the assumption that spacetime is asymptotically flat; see however \cite{Gibbons:1998zr,Chrusciel:2000cu,Chrusciel:2000az,Bhattacharyya:2008xc}.  Instead of attempting to clarify the conditions under which \eno{PenroseBound} must hold, we will make the working assumption that it does hold for the collisions we discuss.

\begin{figure}
  \centerline{\includegraphics[width=5in]{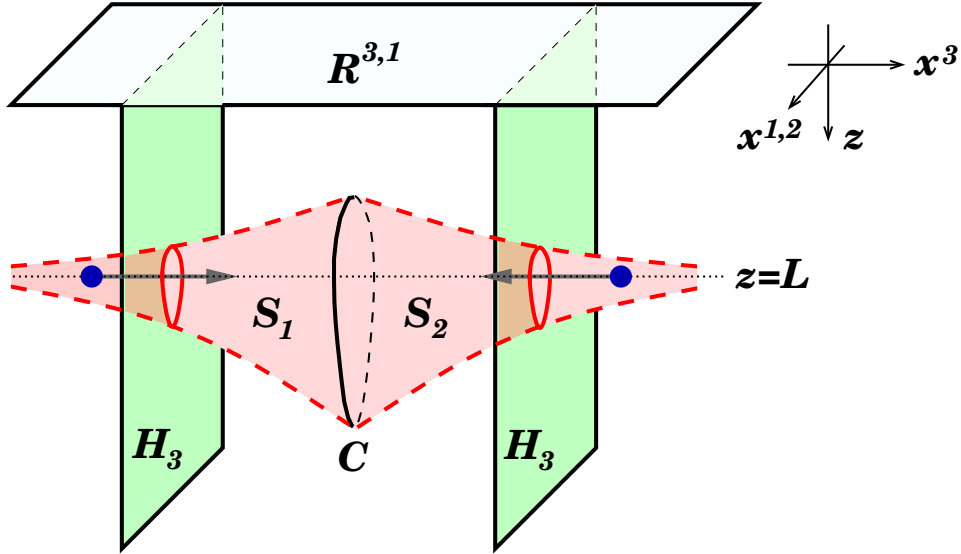}}
  \caption{A projection of the marginally trapped surface that we use onto a fixed time slice of the \AdS geometry.  The size of the trapped surface is controlled by the energy of the massless particles that generate the shock waves.  These particles are shown as dark blue dots.}\label{3Dtrapped}
\end{figure}

When they exist, trapped surfaces are highly non-unique.  But in the case of head-on collisions in flat space there is a standard choice of such surfaces \cite{Penrose,DEath:1992hb,DEath:1992hd,DEath:1992qu} which are easily obtained due to the symmetries of the configuration:
head-on collisions preserve rotational symmetry around the axis of motion of the massless particles, $O(2)$ in $d=4$. The standard trapped surface preserves this symmetry too.  In the case we're considering, the metric \eno{AdSShocks} possesses an $O(3)$ symmetry which acts on $x^1$, $x^2$, and $z$ but preserves $q$.  It is a remnant of the $O(4,2)$ symmetry of \AdS.  We explain this symmetry more fully in section~\ref{SHOCKS}, and in section~\ref{TRAPPED} we construct an $O(3)$-symmetric trapped surface (more precisely, a marginally trapped surface) which is an obvious adaptation of the standard one in flat space.
The marginally trapped surface we find comprises two halves, ${\cal S}_1$ and ${\cal S}_2$, which are matched along a co-dimension three ``curve'' ${\cal C}$. This is depicted in figure~\ref{3Dtrapped}.  ${\cal C}$ lies in a three-dimensional slice of \AdS whose internal geometry is the hyperbolic space $H_3$.  Because of the $O(3)$ symmetry, ${\cal C}$ must be a two-sphere located at some constant value $q_{\cal C}$ of the $O(3)$-invariant variable $q$.

As we will see in section~\ref{TRAPPED}, the area of the trapped surface depends on the energy $E$ of the configuration, and one can obtain a relation between $S_{\rm trapped}$ and $E$ through $q_{\cal C}$.  When $q_{\cal C} \gg 1$, this relation takes the form
 \eqn{TrappedRelations}{
  E \approx {4L^2 \over G_5} q_{\cal C}^3 \qquad
   S_{\rm trapped} \approx {4\pi L^3 \over G_5} q_{\cal C}^2 \,,
 }
from which we can immediately extract
 \eqn{STrappedFromE}{
  S_{\rm trapped} \approx \pi \left( {L^3 \over G_5} \right)^{1/3}
    (2EL)^{2/3} \,.
 }
To obtain a numerical value for $S_{\rm trapped}$, we must evidently select values for the dimensionless quantities $L^3/G_5$ and $EL$.  To choose $L^3/G_5$, consider the translationally invariant \AdS-Schwarzschild solution:
 \eqn{AdSSch}{
  ds^2 = {L^2 \over z^2} \left[ \left( 1-{z^4 \over z_H^4} \right)
    dt^2 + d\vec{x}^2 + {dz^2 \over 1 - {z^4 \over z_H^4}}
     \right] \,.
 }
According to \cite{Gubser:1996de}, the energy density is
 \eqn{AdSSchE}{
  \epsilon = {3\pi^3 \over 16} {L^3 \over G_5} T^4 \,.
 }
On the other hand, lattice calculations\footnote{We took the value quoted in \eno{LatticeEOS} from Figure~1 of \cite{Adcox:2004mh}.  See e.g.~\cite{Karsch:2001cy} for a more comprehensive account.} show that
 \eqn{LatticeEOS}{
  f_* \equiv {\epsilon \over T^4} \approx
    11 \qquad\hbox{for $1.2 T_c < T < 2T_c$}\,,
 }
and that $f_*$ rises slowly above this range.
We choose
 \eqn{FoundL}{
  {L^3 \over G_5} = {16 \over 3\pi^3} \times 11 \approx 1.9
 }
in order to make the black hole equation of state \eno{AdSSchE} match \eno{LatticeEOS}.
Since we have not specified a compact manifold, we need not assume that the \AdS\ background is dual to $SU(N)$ ${\cal N}=4$ super-Yang-Mills.  If we did, \eno{FoundL} would imply that $N \approx 2$.  Instead, we are assuming that the background is an approximate dual to real-world QCD above the confinement transition, or to a theory which is sufficiently close to real-world QCD to make numerical comparisons meaningful.  Alternatively, we are assuming that the dual of the \AdS background captures enough features of real world QCD (above the confinement transition) to make this numerical comparison meaningful.  In any case, loop effects on the gravity side are suppressed only by powers of $G_5/L^3$, so according to \eno{FoundL} they are not very suppressed.  Also, $\alpha'$ corrections could be significant.  Thus, all our calculations are to be understood as leading-order estimates.

To choose a reasonable value of $EL$, we have to know a little more about the holographic dual of a shock wave.  As we explain in section~\ref{SHOCKS}, the expectation value of the gauge theory stress tensor for the right-moving shock is
 \eqn{GotTuu}{
  \langle T_{uu} (\vec{x}) \rangle =
   {L^2 \over 4\pi G_5} \lim_{z\to 0}
   {1\over z^3} \Phi(x^1, x^2, z) \delta(u) =
   {2 L^4 E\over \pi \left(L^2 + (x^1)^2 +
    (x^2)^2\right)^3} \delta(u)\,,
 }
with all other components vanishing when one uses the coordinate system $(u,v,x^1,x^2)$.  Evidently, $E$ is the total energy in the gauge theory.  For gold-gold collisions, $\sqrt{s_{NN}} = 200\,{\rm GeV}$ means $E = E_{\rm beam} = 19.7\,{\rm TeV}$.  $L$ is the rms radius of the transverse energy distribution in \eno{GotTuu}.
Because we're comparing the dual of the shock wave to a boosted gold nucleus, an obvious approach is to set $L$ equal to the rms transverse radius of the nucleons.  Using a Woods-Saxon profile for the nuclear density (see for example \cite{Klein:1999qj,Adams:2004rz}), one obtains an rms transverse radius $L \approx 4.3\,{\rm fm}$.  So we estimate
 \eqn{FoundEL}{
  EL \approx 4.3 \times 10^5 \,.
 }
Putting \eno{STrappedFromE}, \eno{FoundL}, and \eno{FoundEL} together, we find
 \eqn{FoundSofE}{
  S \geq S_{\rm trapped} \approx 35000 \left(
    {\sqrt{s_{NN}} \over 200\,{\rm GeV}} \right)^{2/3} \,.
 }

\begin{figure}
  \centerline{\includegraphics[width=5in]{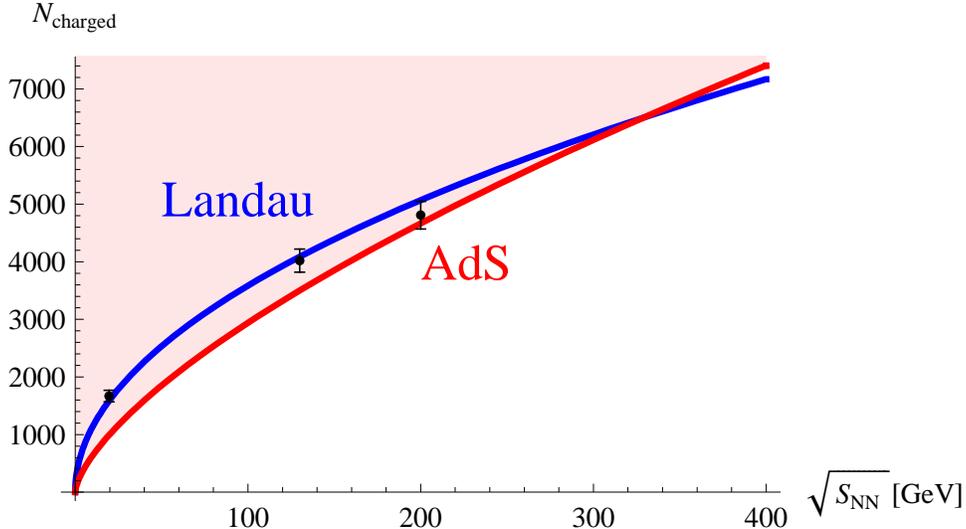}}
  \caption{A plot of the total number of charged particles vs.~energy. The data points were taken from table II of the PHOBOS results \cite{Back:2002wb}.  We show in red the region consistent with the bound \eno{FoundSofE} obtained via the gauge-string duality, using point-sourced shocks and estimates described in the text, and assuming the bound \eno{PenroseBound}.
The blue curve corresponds to the prediction of the Landau model \cite{Landau:1953gs}.}\label{EntropyRHIC}
 \end{figure}

In figure \ref{EntropyRHIC} we have plotted the dependence of the entropy bound \eqref{FoundSofE} on the energy, together with the data from PHOBOS \cite{Back:2002wb}.
It is encouraging that the estimate
\eqref{FoundSofE} for $S_{\rm trapped}$ is just $10\%$ below the phenomenological estimate \eno{Srough} at $\sqrt{s_{NN}} = 200\,{\rm GeV}$.
According to \cite{Dumitru:2007qr}, once we use $\eta/s = 1/{4\pi}$, this is roughly the amount of entropy required to fit a thermalization time of $1$~fm/c.
However, the scaling $S_{\rm trapped} \propto s_{NN}^{1/3}$ implied by \eqref{FoundSofE} differs from the observed scaling, which is closer to the dependence $S \propto s_{NN}^{1/4}$.
As observed in \cite{Carruthers:1973dw}, the latter dependence, predicted by the Landau model \cite{Landau:1953gs},\footnote{For an introduction to the Landau model, see section~\ref{LANDAU}; for a review, see \cite{Steinberg:2004vy}.} seems to hold over a strikingly large range of energies. Put differently, the inequality in \eno{FoundSofE} is consistent with all heavy-ion collision data to date, but for energies only slightly above RHIC energies, \eno{FoundSofE} predicts a faster increase of entropy than is generally expected.

At the LHC, $\sqrt{s_{NN}}$ will be $5.5\,{\rm TeV}$ for lead-lead collisions.  Inserting this value into \eno{FoundSofE}, and making minor corrections for the differences between lead and gold\footnote{$A=208$ for lead, so $E_{\rm beam} = 570\,{\rm TeV}$; $L=4.4\,{\rm fm}$ from the rms radius of lead, resulting in $EL \approx 1.3 \times 10^7$.} one finds
 \eqn{LHCmultiplicity}{
  S_{\rm trapped} \approx 3.4 \times 10^5\,.
 }
$S \geq S_{\rm trapped}$ corresponds to $N_{\rm charged} \geq 45000$ if we continue to use \eno{Srough}.
The lower bound on the entropy \eno{LHCmultiplicity} exceeds the prediction of the Landau model, $S \approx 2.1 \times 10^5$, by a factor of about $1.6$.  Calculations based on the Color Glass Condensate tend to predict lower multiplicities: for example, from figure~5 of \cite{Kharzeev:2004if}, one may read off the prediction $N_{\rm charged} \approx 22000$, about a factor of $2$ below the estimate from \eno{LHCmultiplicity}; see also \cite{Abreu:2007kv}.

We see three main ways in which \eno{FoundSofE} could fail:
 \begin{enumerate}
  \item Using the gauge-string duality to describe entropy production may cause us to misrepresent perturbative aspects of the early stages of the collision.  This is because our use of the gauge-string duality relies on the supergravity approximation, which is the leading order description of a strong coupling expansion as well as a $1/N$ expansion.  Our methods appear to offer no access to perturbative physics.  Perturbative QCD is expected to characterize the early stages of LHC collisions more cleanly than it does RHIC collisions, and it may be that our methods are correspondingly less applicable at LHC than at RHIC.
  \item As we will see in section~\ref{MINKOWSKI}, there is a whole family of \AdS shock waves with the same $\langle T_{uu} \rangle$, presumably distinguished by higher point functions of $T_{uu}$.  The trapped surface depends on which of these shock waves we pick.  It is easy to lower $S_{\rm trapped}$ by spreading the shock wave out over the transverse $H_3$ in \AdS.  So although \eno{FoundSofE} at first looks highly predictive, and easily falsifiable at energies significantly higher than RHIC scales, it is in fact possible to accommodate slower growth of total entropy with beam energy.
We discuss this further in section \ref{DISCUSSION}.
  \item The bound \eno{PenroseBound} could fail, even for standard Einstein gravity in \AdS.
 \end{enumerate}

The rest of this paper is organized as follows.  In section~\ref{PHENOMENOLOGY}, we review phenomenological estimates of the entropy produced in a heavy ion collision, with the aim of justifying \eno{Srough} and briefly summarizing the dependence on beam energy.  In section~\ref{SHOCKS}, we review the construction of shock waves in AdS${}_D$, with particular attention to the $O(D-1)$ symmetry preserved by head-on collisions of the simplest shock wave constructions.
These symmetries might be approximately realized in central heavy-ion collisions even if gauge-string methods fail to give a quantitatively accurate account of entropy production.  Aside from discussing these symmetries, our purpose in sections~\ref{PHENOMENOLOGY} and~\ref{SHOCKS} is mostly to gather together well-known facts from the literature.  Our main calculations are in section~\ref{TRAPPED}, where we compute the shape of marginally trapped surfaces.  We end with a discussion in section~\ref{DISCUSSION}.

\section{Phenomenological estimates of the entropy}
\label{PHENOMENOLOGY}

In order to evaluate the entropy $S$ produced in a heavy-ion collision, one needs a method to relate the entropy to a quantity which can be measured: the number of charged particles, $N_{\rm charged}$. In~\ref{BJORKEN} and~\ref{PHASE} we review two such methods. The first uses the framework of Bjorken flow \cite{Bjorken:1982qr}. The other, described in section~\ref{PHASE}, relies on phase space estimates to evaluate the ratio $S/N_{\rm charged}$ after hadronization.  Both of these sections largely follow \cite{Pal:2003rz}.

While sections~\ref{BJORKEN} and~\ref{PHASE} allow an evaluation of the entropy via the measured number of charged particles, in section~\ref{LANDAU} we estimate the entropy from the size and shape of the colliding nuclei---or, more precisely, the size and shape of the parts of the nuclei that participate significantly in the collision, and the beam energy per nucleon. This last estimate is based on the Landau model \cite{Landau:1953gs}.

Non-specialists may appreciate the reminder that $s_{NN}$ is the Mandelstam variable for a pair of nucleons, one from each nucleus.  When the beam energy is $100\,{\rm GeV}$ per nucleon, $\sqrt{s_{NN}} = 200\,{\rm GeV}$\@.  Because gold has $197$ nucleons, the total center of mass energy is $39.4\,{\rm TeV}$\@.  It is also good to know that the rapidity of a particle emerging from the collision is $y = \tanh^{-1} p_z/E$, whereas pseudo-rapidity is $\eta = \tanh^{-1} p_z/p = \tanh^{-1} \cos\theta$, where $\theta$ is the angle from the beamline.

\subsection{Entropy estimate from Bjorken flow}
\label{BJORKEN}
In this section (as well as in parts of section~\ref{PHASE}) we estimate the total entropy produced in the collision by assuming that the entropy per charged particle changes only slightly with rapidity: thus
 \eqn{TotalMid}{
  {S \over N_{\rm charged}} \approx \left.
    {dS/dy \over dN_{\rm charged}/dy} \right|_{\rm mid-rapidity} \,.
 }
Both $N_{\rm charged}$ and $dN_{\rm charged}/dy$ at mid-rapidity are directly measured, so we only need to estimate $dS/dy$ at mid-rapidity.  To do this, we follow \cite{Pal:2003rz} and consider Bjorken's boost-invariant treatment of a collision.

One of the main relations emerging from Bjorken's treatment is
 \eqn{BjorkenEnergy}{
  \tau_{\rm form} A \, \epsilon(\tau_{\rm form}) =
    {dE_T \over dy} \,,
 }
where $\epsilon$ is the energy density, $A$ is the cross-sectional area of the participating nucleons, $\tau_{\rm form}$ is the formation time, and $E_T$ is the transverse energy of a particle, defined as $E\sin\theta$ where $E$ is the total energy and $\theta$ is the angle from the beamline.\footnote{Sometimes the definition of $E_T$ is varied across particle species by adding some multiple of the rest mass: see for example \cite{Adams:2004cb}.}
The entropy may be expected to follow a similar relation:
 \eqn{BjorkenEntropy}{
  \tau_{\rm form} A \, s(\tau_{\rm form}) = {dS \over dy} \,.
 }
If the QGP is a thermalized plasma at a time $\tau_{\rm form}$ (which may not be true but provides a rough estimate), then assuming conformal invariance one has
 \eqn{ConformalPlasma}{
  s = {4 \over 3} {\epsilon \over T} \,.
 }
By solving \eno{BjorkenEnergy} for $\epsilon(\tau_{\rm form})$ and \eno{BjorkenEntropy} for $s(\tau_{\rm form})$, and then plugging the resulting expressions into \eno{ConformalPlasma}, one arrives at
 \eqn{EntropyDensity}{
  {dS \over dy} = {4 \over 3T} {dE_T \over dy} \,.
 }
The quantities in \eno{EntropyDensity} are all to be evaluated at $\tau_{\rm form}$, but for a rough estimate one may use
 \eqn{ETformBound}{
  {dE_T(\tau_{\rm form}) \over dy} \approx
    {dE_T({\rm final}) \over d\eta} \approx 600\,{\rm GeV}
 }
for central gold-gold collisions at $\sqrt{s_{NN}} = 200\,{\rm GeV}$ \cite{Adcox:2004mh}.\footnote{$\sqrt{s_{\rm NN}} = 200\,{\rm GeV}$ means that a pair of nucleons, one from each nucleus, together have Mandelstam $s = (200\,{\rm GeV})^2$.  That means the beam energy is $100\,{\rm GeV}$ per nucleon, and because gold has $197$ nucleons, the total center of mass energy is $39\,{\rm TeV}$.}
To find the temperature, one may resort again to \eno{BjorkenEnergy} together with the approximation \eno{LatticeEOS} of the lattice equation of state.  Choosing the nominal values $\tau_{\rm form} = 1\,{\rm fm}$ and $A = 120\,{\rm fm}^2$ leads to $T = 240\,{\rm MeV}$.  Then \eno{EntropyDensity} together with \eno{ETformBound} give
 \eqn{EntropyBound}{
  {dS(\tau_{\rm form}) \over dy} \approx 3300 \,.
 }

The number of charged tracks per unit rapidity for a central collision is
 \eqn{ChargedTracks}{
  {dN_{\rm charged} \over dy} \approx 660
 }
near mid-rapidity (see for example \cite{Back:2004je}).  Once again as a rough estimate, let's use
 \eqn{SRoughFormula}{
  {dS({\rm final}) \over dy} \approx {dS(\tau_{\rm form}) \over dy}
    \,.
 }
Putting \eno{EntropyBound}--\eno{SRoughFormula} together, we arrive at
 \eqn{EntropyTracks}{
  {dS(\tau_{\rm form}) \over dy} \approx
    5 {dN_{\rm charged} \over dy} \,.
 }
Using \eno{TotalMid} leads to
 \eqn{TotalEntropy}{
  S \approx 5 N_{\rm charged} \approx 25000 \,,
 }
where we recalled that $N_{\rm charged} \approx 5000$ in a central collision.  It's important to bear in mind that many approximations were used in arriving at \eno{TotalMid}, so it should be regarded only as a first attempt.  Some refinements were outlined in \cite{Pal:2003rz}.

\subsection{Entropy estimates from phase space density}
\label{PHASE}

Phase space estimates of entropy start from the expressions
 \eqn{SandNphase}{
  S &= \sum_i \int {d^3 x \, d^3 p \over (2\pi)^3}
   \left[ -f_i \log f_i + s_i (1+s_i f_i) \log(1+s_i f_i) \right]
     \cr
  N &= \sum_i \int {d^3 x \, d^3 p \over (2\pi)^3} f_i \,,
 }
where $f_i = f_i(x,p)$ is the phase space density for each spin polarization of each hadronic species, and $s_i=1$ for bosons and $-1$ for fermions.  The number of hadrons $N$ is roughly ${3 \over 2} N_{\rm charged}$.  Using \eno{SandNphase} means that one is ignoring interactions among hadrons, so it should apply, in some approximation, after hadronization.  One line of thought \cite{Sollfrank:1992ru,Nonaka:2005vr,Muller:2005en} is to use the equilibrium expressions
 \eqn{fEquil}{
  f_i = {1 \over e^{\sqrt{p^2+m_i^2}/T} - s_i} \,,
 }
set $T=170\,{\rm MeV}$, and run the sums in \eno{SandNphase} over all established hadron resonances.  The result is
 \eqn{fSoverN}{
  {S / N} = 5.15 \,.
 }
Applying \eno{fSoverN} to the total entropy in a heavy ion collision gives
 \eqn{fSoverNch}{
  {S / N_{\rm charged}} = 7.7 \,.
 }
The estimate \eno{fSoverNch} is at best approximate, because chemical potentials for quarks become significant at forward rapidities. Replacing $S$ and $N_{\rm charged}$ by $dS/dy$ and $dN_{\rm charged}/dy$ in \eno{fSoverNch} would improve the reliability of the estimate.  But \eno{fSoverNch} is also approximate because the system is not really an equilibrated gas of nearly free hadrons at $T=170\,{\rm MeV}$; rather, it is at roughly this temperature that the quark-gluon plasma hadronizes.

A more data-driven approach was taken in \cite{Pal:2003rz}: instead of assuming \eno{fEquil}, experimental results for single-particle yields and two-particle interferometry were used to estimate the $f_i$.  For central collisions, and at mid-rapidity, one finds from this approach the result
 \eqn{dSdyPratt}{
  {dS \over dy} = 4451
    \qquad\hbox{at $\sqrt{s_{NN}} = 130\,{\rm GeV}$\,.}
 }
Combining \eno{dSdyPratt} with
 \eqn{dNchargedPratt}{
  {dN_{\rm charged} \over dy} \approx 620
    \qquad\hbox{at $\sqrt{s_{NN}} = 130\,{\rm GeV}$}
 }
(see for example \cite{Back:2002wb}) gives the ratio\footnote{The STAR collaboration has published a result corresponding to $dN_{\rm charged}/dy \approx 580$, which would result in a value $7.7$ in \eno{SoverNPratt}.  An even higher value, $8.5$, can be read off from estimates in \cite{Nonaka:2005vr}; however there seems to be some possible confusion about $dN_{\rm charged}/dy$ versus $dN_{\rm charged}/d\eta$.
We have used a common mid-rapidity conversion factor, $dN_{\rm charged}/dy \approx 1.1 dN_{\rm charged}/d\eta$, to pass from results quoted in terms of pseudo-rapidity densities to rapidity densities.}
 \eqn{SoverNPratt}{
  {dS / dy \over dN_{\rm charged} / dy} = 7.2 \,.
 }
We arrived at the figure \eno{Srough} simply by using $S/N_{\rm charged} = 7.5$, an average of \eno{fSoverNch} and~\eno{SoverNPratt}.

\subsection{Entropy estimates from immediate equilibration}
\label{LANDAU}

The Landau model of particle production in high-energy collisions \cite{Landau:1953gs} assumes that hydrodynamics is valid starting from the moment that the colliding nuclei completely overlap.  It also assumes a conformal equation of state, $p=\epsilon/3$.  The validity of hydrodynamics depends on local thermodynamic equilibrium and a mean free path that is short compared to the extent of the medium.  Total overlap occurs about $0.13\,{\rm fm}$ after the nuclei first start to interact in a central gold-gold collision at $\sqrt{s_{NN}} = 200\,{\rm GeV}$.  It doesn't seem reasonable to assume that hydro is valid at such an early time, so it is surprising how well the model works in describing aspects of the bulk flow, in particular the particle distribution in rapidity.

The entropy is easy to estimate at $\tau_{\rm overlap}=0.13\,{\rm fm}$.  We should ignore the nucleons which do not interact: this includes a good fraction of the ones in the outer skin, or corona, of the gold nucleus.  In a central collision (more precisely, in the $5\%$ of collisions that are the most central) a typical number of participating nucleons is $N_{\rm part} = 350$, so the total energy of participating nuclei is
 \eqn{Eparticipating}{
  E_{\rm tot} =
    {N_{\rm part} \sqrt{s_{NN}} \over 2} = 35\,{\rm TeV} \,.
 }
In the rest frame of one nucleus, its participants occupy a roughly spherical region of radius $6.5\,{\rm fm}$, which we will assume to have uniform density.  Let's denote the volume of this sphere by $V$.  In the lab frame, this sphere is Lorentz flattened by a factor $\gamma = \sqrt{s_{NN}}/2m_p$, where $m_p = 0.938\,{\rm GeV}$ is the mass of a proton.  Thus the energy density at the moment of overlap is
 \eqn{EnergyDensityLandau}{
  \epsilon = {\gamma E_{\rm tot} \over V}
    = 3300 \, {\rm GeV}/{\rm fm}^3 \,.
 }
Using \eno{LatticeEOS}, the corresponding temperature is
 \eqn{TemperatureLandau}{
  T = (\epsilon/f_*)^{1/4} = 1200 \, {\rm MeV} \,,
 }
and the entropy is
 \eqn{EntropyLandau}{
  S = {4 \over 3} {E_{\rm tot} \over T}
    = {2 \over 3} (4 f_* m_p V N_{\rm part}^3
       s_{NN})^{1/4} \approx 38000 \,,
 }
which is fortuitously close to \eno{Srough}.  If one started instead by assuming that all the nucleons participate and that the radius is $7\,{\rm fm}$, the entropy estimate would increase to $44000$.  The usual assumption in the Landau model is that subsequent expansion is isentropic.

To arrive at the figure $S \approx 2.1 \times 10^5$ for entropy production at the LHC, quoted below \eno{LHCmultiplicity}, we used \eno{EntropyLandau} with $N_{\rm part} = 368$ (scaled up from the number for gold in linear proportion to the atomic number), $R = 6.6\,{\rm fm}$ (scaled up from the number for gold in proportion to the cube root of the atomic number), and the same value $f_*=11$ as before.\footnote{A fractionally higher value, say $f_*=12$, might be closer to lattice values, but it doesn't make a difference at the level of accuracy we have quoted.}

\section{Shock waves in anti-de Sitter space}
\label{SHOCKS}

Gravitational shock waves are well studied, both in ${\bf R}^{D-1,1}$ and AdS${}_D$: see for example \cite{Aichelburg:1970dh, Dray:1984ha, Hotta:1992qy, Sfetsos:1994xa, Podolsky:1997ni, Horowitz:1999gf, Emparan:2001ce, Arcioni:2001my, Kang:2004jd, Cornalba:2006xk}.  The simplest of them can be constructed in two equivalent ways.
One is to boost a black hole in AdS${}_D$ to a velocity approaching the speed of light, while at the same time decreasing the mass of the black hole in such a way that the energy remains fixed.  We describe this construction in section~\ref{BOOSTING} for the special case $D=5$.  Alternatively, one can start off with a pointlike, massless particle traveling in AdS${}_D$ and show that it back-reacts on the metric in such a way as to produce a shock-wave discontinuity.  We describe this construction for arbitrary $D$ in section~\ref{DIMENSIONS}.
Other types of shocks obtained by sourcing the metric with appropriate matter are given in section \ref{OTHER}.  Our goal is to relate collisions of shock waves to collisions of heavy ions, and to this end it is useful to have the energy density dual to the colliding shocks. We compute this in section \ref{MINKOWSKI}.

\subsection{Constructing the simplest shock wave geometry}
\label{BOOSTING}

Our starting point is the global \AdS-Schwarzschild (GAdSBH) metric,
 \eqn{GAdSBH}{
   ds^2 = -f d\tau^2+ {d\rho^2\over f} + \rho^2 d \Omega_3^2 \qquad
    f \equiv 1 + {\rho^2\over L^2} - {\rho_0^2\over \rho^2}\,,
 }
where the parameter $\rho_0$ can be related to the ADM mass of the black hole by
 \eqn{ADMMass}{
  M = {3 \pi \over 8G_5} \rho_0^2\,.
 }
Since we are working in global coordinates, the boundary theory has topology $S^3 \times {\bf R}$.  Working in a coordinate system which covers only the Poincar\'e wedge of AdS corresponds to putting the boundary theory on ${\bf R}^{3,1}$.

To make this more precise, recall that \AdS is the universal cover of the five-dimensional hyperboloid
 \eqn{AdSHyperb}{
  -(X^{-1})^2 - (X^0)^2 + (X^1)^2 + (X^2)^2 + (X^3)^2 + (X^4)^2 = -L^2
 }
in ${\bf R}^{4,2}$.  The metric of \AdS, which is given by \eqref{GAdSBH} with $\rho_0=0$, is also the metric induced on the hyperboloid from the standard flat metric on ${\bf R}^{4,2}$.  The $X^M$ coordinates are related to the global coordinates $(\tau, \rho, \Omega^i)$ in \eqref{GAdSBH} as follows:
 \eqn{GlobalToX}{
  X^{-1} = \sqrt{\rho^2 + L^2} \cos{\tau\over L} \qquad X^0 = \sqrt{\rho^2 + L^2} \sin{\tau\over L} \qquad X^i = \rho \Omega^i\,,
 }
where $\Omega^i$ is a unit vector in ${\bf R}^4$, which is to say a point on $S^3$. But $\tau$ runs from $-\infty$ to $\infty$ in \AdS, whereas $\tau=0$ and $\tau=2\pi L$ are identified on the hyperboloid.
Thus, the coordinates $X^M$ are more fit to describe the hyperboloid of which \AdS is the covering space, while in the $(\tau,\rho,\Omega^i)$ coordinate system, \AdS can be thought of as a cylinder with boundary $S^3 \times \mathbf{R}$ if we conformally compactify in the $\rho$ direction. A more detailed discussion can be found, for example, in \cite{Aharony:1999ti}.
The Poincar\'e coordinates $(t, x^1, x^2, x^3, z)$
are related to the $X^M$ coordinates by
 \eqn{GlobalToPoincare}{
  X^{-1} = {z\over 2} \left(1+ {L^2 + \vec{x}^2 -t^2\over z^2}\right)
   \qquad X^0 = L {t\over z}\cr
  X^i = L{x^i\over z} \qquad X^4 =
   {z\over 2} \left(-1 + {L^2 - \vec{x}^2 + t^2\over z^2}\right)\,.
 }
The actual metric obtained when transforming \eqref{GAdSBH} to the Poincar\'e patch is somewhat complicated, and we shall not write it explicitly here.

\begin{figure}
  \centerline{\includegraphics[width=5in]{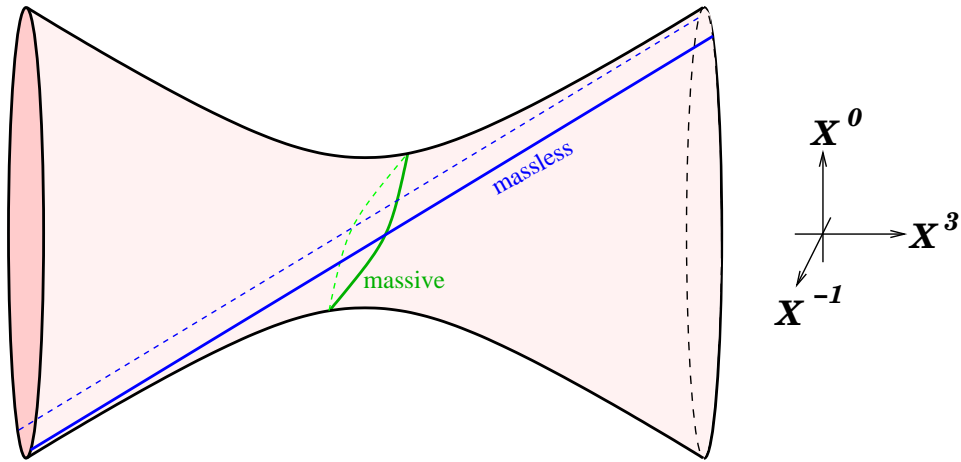}}
  \caption{The hyperboloid whose covering space is \AdS, with the transverse coordinates $X^1$, $X^2$, and $X^4$ suppressed.  The closed green curve is the trajectory of a massive test particle.  When the particle is infinitely boosted, so that $X^0=X^3$, the trajectory degenerates into the two blue lines.}\label{hyperboloid}
 \end{figure}

We wish to boost this black hole to the speed of light while decreasing its mass and keeping its energy constant.  The method we use is similar to the one explained in \cite{Hotta:1992qy}.  Following \cite{Aichelburg:1970dh} we expect that this boost will give us the gravitational field around a massless particle moving in \AdS. Thus, we choose a boost that will take a stationary particle to one moving at a highly relativistic speed. A massive test particle follows a closed trajectory which is described, in terms of the $X^M$ coordinates, by the intersection of the hyperboloid \eqref{AdSHyperb} with the plane $X^3=\beta X^0$ with fixed $X^1$, $X^2$, and $X^4$.  A convenient choice is $X^1=X^2=X^4=0$.  In figure \ref{hyperboloid} we have shown one such trajectory.
Once we take the mass of the particle to zero, the trajectory of the test particle degenerates into two straight lines (straight both in the sense of being geodesics on the hyperboloid and in the sense of the flat metric of ${\bf R}^{4,2}$). This is also depicted in figure~\ref{hyperboloid}.
In global coordinates, a massive particle starting at $\rho=0$ with some non-zero velocity returns to $\rho=0$ with the opposite velocity after a time $\tau = \pi L$, then continues to oscillate through the \AdS cylinder with period $\tau = 2\pi L$.
When taking the lightlike limit of the trajectory, the oscillating motion of the massive particle deforms into a bouncing motion, going from one boundary to the other. Each leg takes a finite global time $\Delta\tau = \pi L$ and corresponds to one line on the hyperboloid.

To carry out the boost explicitly we note that the isometry of the hyperboloid \eno{AdSHyperb} is the $O(4,2)$ group of linear transformations preserving the quadratic form on the left hand side of \eno{AdSHyperb}.  The boost that we'll be interested in is an element of the $SO(1,1)$ subgroup which preserves $X^{-1}$, $X^1$, $X^2$, $X^4$, and the quadratic form $-(X^0)^2 + (X^3)^2$:
 \eqn{Boost}{
  X^0 \to \tilde{X}^0 \equiv {X^0 - \beta X^3\over \sqrt{1-\beta^2}} \qquad X^3 \to \tilde{X}^3 \equiv {-\beta X^0 + X^3\over \sqrt{1-\beta^2}}\,.
 }
Writing $M = E \sqrt{1-\beta^2}$ and taking $\beta\to 1$ with $E$ held fixed, the boosted GAdSBH metric \eqref{GAdSBH} becomes
 \eqn{AlmostShock}{
  ds^2 &= ds_{AdS_5}^2 + {8 G_5 L^2 E  \over 3  \beta \sqrt{1-\beta^2}} { \left[(\tilde{X}^0)^2-(X^{-1})^2 \right] L^2 + \left[(\tilde{X}^0)^2+(X^{-1})^2 \right] (X^{-1})^2 \over \left[(\tilde{X}^0)^2+(X^{-1})^2 \right] \left[-L^2 + (\tilde{X}^0)^2+(X^{-1})^2 \right]}(dX^0 - dX^3)^2 \cr
  &{}+ {\cal O}\left(\sqrt{1-\beta^2}\right)\,.
 }
Using
 \eqn{LargeVLimit}{
  \lim_{\beta\to 1} {1\over \sqrt{1-\beta^2}} f\left(X^0 - \beta X^3\over \sqrt{1-\beta^2}\right) = \delta(X^0 - X^3) \int_{-\infty}^{\infty} f(x) dx
 }
which holds for any integrable function $f$, the $\beta\to 1$ limit of \eqref{AlmostShock} becomes
 \eqn{Shock}{
  ds^2 &= ds_{AdS_5}^2 - {4 G_5 E  \left[L^2 - 2 (X^{-1})^2 + 2 X^{-1} \sqrt{(X^{-1})^2-L^2} \right] \over L^2 \sqrt{(X^{-1})^2-L^2}}  \cr
    &\qquad\qquad{} \times \Theta(X^{-1} - L) \delta(X^0 - X^3) (dX^0 - dX^3)^2\,,
 }
where
 \eqn{thetaDef}{
  \Theta(x) \equiv \left\{ \eqalign{1 &\qquad\hbox{if $x \geq 0$}  \cr
    0 &\qquad\hbox{if $x < 0$.}} \right.
 }
In Poincar\'e coordinates \eqref{GlobalToPoincare}, the line element \eqref{Shock} reads
 \eqn{ShockPoincare}{
  ds^2 = {L^2 \over z^2} \left[ -du dv +
    (dx^1)^2 + (dx^2)^2 + dz^2 \right] +
     {L\over z} \Phi(x^1, x^2, z) \delta(u) du^2 \,,
 }
where $u$, $v$, and $\Phi$ are defined as in \eno{uvDef}--\eno{PhiDef}.  In checking the equivalence of \eno{Shock} and \eno{ShockPoincare}, it helps to note that
 \eqn{qXrelation}{
  q = {X^{-1} - L \over 2L} \,.
 }

A subtlety in the derivation above is the emergence of the factor of $\Theta(X^{-1}-L)$. As we've explained earlier, the massless trajectory comprises of two straight lines describing a massless particle which goes back and forth from one boundary of the AdS cylinder to the other. We choose to consider only one such leg, and this is what the factor $\Theta(X^{-1}-L)$ does.  Including the return journey of the particle would correspond to adding an additional term to \eno{Shock} identical to the one explicitly shown, but with $X^{-1} \to -X^{-1}$.

The metric \eno{ShockPoincare} has an $O(3)$ symmetry which is simpler to understand in the $X^M$ coordinates:  the boosted metric \eno{AlmostShock} doesn't depend on $X^1$, $X^2$, or $X^4$, except through the constraint \eno{AdSHyperb}, which (after the boost) can be regarded as a way to determine $X^{-1}$ in terms of $\tilde{X}^0$, $\tilde{X}^3$, and $(X^1)^2+(X^2)^2+(X^4)^2$.  The $O(3)$ symmetry is the one acting on the coordinates $X^1$, $X^2$, and $X^4$ transverse to the particle's trajectory. To see the transverse space more clearly, we slice \AdS at a definite value of $X^0$ and impose $X^0=X^3$. This gives a two-sheeted hyperboloid:
 \eqn{TwoSheeted}{
  -(X^{-1})^2 + (X^1)^2 + (X^2)^2 + (X^4)^2 = -L^2 \,.
 }
The two disjoint sheets are related by $X^{-1} \to -X^{-1}$.
Each is a copy of the Euclidean hyperbolic space $H_3$.  The massless particle that we're interested in passes through the ``center'' of the upper sheet, at $X^{-1}=L$ and $X^1=X^2=X^4=0$.  (We write ``center'' in quotes because $H_3$ is a homogeneous space.)  The isometries of $H_3$ form $O(3,1)/{\bf Z}_2$, and the $O(3)$ of interest is the part of this group that preserves the point that the massless particle passes through.
Note that the $O(3)$ symmetry we've found is not equivalent to rotational symmetry in the $(x^1,x^2,x^3)$ plane of the Poincar\'e patch. Rather, because it acts non-trivially on $X^4$, its generators include special conformal transformations.

\subsection{Shock wave metrics in AdS${}_D$}
\label{DIMENSIONS}

The metric \eqref{ShockPoincare} can also be obtained by solving the Einstein equations in the presence of a lightlike particle.  This alternative derivation is a little more efficient, and we will take advantage of this to generalize to $D$ dimensions.  Following \cite{Dray:1984ha,Kang:2004yk,Nastase:2004pc}, one starts with an ansatz
 \eqn{PoincareAnsatz}{
  ds^2 = {L^2 \over z^2} \left( -du dv +
    \sum_{i=1}^{D-3} (dx^i)^2 + dz^2 +
     \phi(x^i,z) \delta(u) du^2 \right) \,.
 }
If $\phi=0$, this is the metric of AdS${}_D$ in Poincar\'e coordinates.  Because \eno{PoincareAnsatz} is supposed to be the metric in the presence of matter, it should satisfy the appropriate Einstein equations:
 \eqn{E:Einstein}{
  R_{\mu\nu} - {1 \over 2} g_{\mu\nu} R -
    {(D-1)(D-2) \over 2 L^2} g_{\mu\nu} = 8 \pi G_D J_{\mu\nu} \,,
 }
where $J_{\mu\nu}$ is the bulk stress tensor, not to be confused with the boundary stress tensor $T^{mn}$.  For a massless particle with energy $E$, the only non-zero component of $J_{\mu\nu}$ is
 \eqn{Jmn}{
  J_{uu} = E \, \delta(u) \delta(z-L) \prod_{i=1}^{D-3}
    \delta(x^i) \,.
 }
It is straightforward to plug \eno{PoincareAnsatz} and \eno{Jmn} into the $uu$ component of \eno{E:Einstein} and explicitly derive
 \eqn{LaplaceShock}{
  \left( \square_{H_{D-2}} - {D-2 \over L^2} \right) \Phi =
   -16\pi G_D \, E \delta(z-L) \prod_{i=1}^{D-3}
     \delta(x^i) \,,
 }
where
 \eqn{PhiTransition}{
  \Phi = {L \over z} \phi
 }
and
 \eqn{squareD}{
  \square_{H_{D-2}} = {z^{D-2} \over L^2} {\partial \over \partial z}
     z^{4-D} {\partial \over \partial z} +
    {z^2 \over L^2} \sum_{i=1}^{D-3} \left(
      {\partial \over \partial x^i} \right)^2
 }
is the laplacian on the Euclidean hyperbolic space $H_{D-2}$, whose line element is
 \eqn{dsHD}{
  ds_{H_{D-2}}^2 = {L^2 \over z^2} \left( \sum_{i=1}^{D-2}
    (dx^i)^2 + dz^2 \right) \,.
 }
Evidently, $H_{D-2}$ is the space transverse to the trajectory of a massless particle.  If we introduced global coordinates $X^M$ on AdS${}_D$, it would have a description entirely analogous to the one explained around \eno{TwoSheeted}.  For our present purposes, it is enough to introduce a subset of the global coordinates, as follows:
 \eqn[c]{GlobalToPoincareY}{
  Y^0 = {z\over 2} \left( 1 + {L^2 + \sum_{i=1}^{D-3}(x^i)^2 \over z^2}
    \right) \qquad
  Y^{D-2} = {z\over 2} \left( -1 + {L^2 - \sum_{i=1}^{D-3}(x^i)^2 \over z^2}
    \right)  \cr
  Y^i = {L \over z} x^i \qquad\hbox{for $i=1$ through $D-3$.}
 }
$H_{D-2}$ is the upper sheet of the two-sheeted hyperboloid
 \eqn{H3Hyperboloid}{
  -(Y^0)^2 + \sum_{i=1}^{D-2} (Y^i)^2 = -L^2 \,,
 }
which has isometry group $O(D-2,1)/{\bf Z}_2$.\footnote{The isometry group of the full two-sheeted hyperboloid is $O(D-2,1)$.  The ${\bf Z}_2$ that one must divide out when considering a single sheet acts by sending $Y^0 \to -Y^0$.} In analogy to \eno{GlobalToX} one may define
 \eqn{GlobalToY}{
  Y^0 = \sqrt{L^2+r^2} \qquad
  Y^i = r\Omega^i \qquad\hbox{for $i=1$ through $D-2$,}
 }
where $\Omega^i$ is a point on a unit $S^{D-3}$.  The metric \eno{dsHD} can be re-expressed as
 \eqn{dsHDagain}{
  ds_{H_{D-2}}^2 = {dr^2 \over 1+r^2/L^2} + r^2 d\Omega_{D-3}^2 \,.
 }
We will be especially interested in the quantity
 \eqn{qDefAgain}{
  q \equiv {\sum_{i=1}^{D-3} (x^i)^2 + (z-L)^2 \over 4zL}
    = {1 \over 4L^2} \left(
   -(Y^0-L)^2 + \sum_{i=1}^{D-2} (Y^i)^2 \right) \,,
 }
where the second equality can be checked using \eno{GlobalToPoincare}.  The last expression in \eno{qDefAgain} shows that, up to an overall prefactor, $q$ is the square of the chordal distance between the point $(x^i,z)$ on $H_{D-2}$ and the special point $(x_*^i,z_*) = (\vec{0},L)$ through which the massless particle passes.  Chordal distance, by definition, is the distance in the embedding space ${\bf R}^{D-2,1}$ parameterized by the coordinates $(Y^0,Y^i)$.  In the coordinates introduced in \eno{GlobalToY}, the special point is at $r=0$, and
 \eqn{ChordalToR}{
  q = {-1 + \sqrt{1+r^2/L^2} \over 2} \qquad
   r = 2L \sqrt{q(1+q)} \,.
 }
Thus one may re-express
 \eqn{MetricChordal}{
  ds_{H_{D-2}}^2 = L^2 \left[
    {dq^2 \over q(1+q)} + 4q(1+q) d\Omega_{D-3}^2 \right] \,.
 }
The $O(D-2)$-symmetric solutions to \eno{LaplaceShock} can be efficiently found by rewriting it in terms of $q$:
\eqn{LaplaceShockAgain}{
  q(1+q)\Phi^{\prime\prime}+\frac{1}{2}(1+2q)(D-2)\Phi^{\prime}-(D-2)\Phi = -{2^{7-D} \pi G_D E \over L^{D-4} (\Vol S^{D-3})}
     {\delta(q) \over [q(1+q)]^{(D-4)/2}} \,,
 }
where
 \eqn{VolSphere}{
  \Vol S^{D-3} = {(D-2)\pi^{(D-2)/2} \over \Gamma(D/2)} \,.
 }
The solution to \eno{LaplaceShockAgain} with the boundary condition that $\Phi(q) \to 0$ as $q\to \infty$ is
 \eqn{PhiGeneralSoln}{
  \Phi(q) = {2^{6-D} \pi G_D E \over L^{D-4} (\Vol S^{D-3})}
    {q^{2-D} \over D-1} \,
    {}_2F_1(D-2,D/2;D;-1/q) \,.
 }
It is easy to check that \eno{PhiDef} is recovered by setting $D=5$.

\subsection{Other sources as shock waves}
\label{OTHER}
Although gravitational shock waves are solutions of the full non-linear Einstein equations (in a distributional sense), two shocks moving in the same direction can be superposed: that is, if one starts with
 \eqn{FirstShock}{
  ds^2 = ds_{AdS_D}^2 + {L \over z} \Phi_1(x^i,z) \delta(u-u_1) du^2
 }
as the first shock and
 \eqn{SecondShock}{
  ds^2 = ds_{AdS_D}^2 + {L \over z} \Phi_2(x^i,z) \delta(u-u_2) du^2
 }
as the second, then the superposed solution is
 \eqn{SuperposedShocks}{
  ds^2 = ds_{AdS_D}^2 + {L \over z} \left[ \Phi_1(x^i,z)
    \delta(u-u_1) + \Phi_2(x^i,z) \delta(u-u_2) \right] du^2 \,.
 }
If \eno{FirstShock} and \eno{SecondShock} are sourced by massless point particles, then \eno{SuperposedShocks} describes the back-reaction of the two massless particles together.  More generally, we can consider a cloud of massless particles, all moving in the same direction, and then the metric is
 \eqn{CloudMetric}{
  ds^2 = ds_{AdS_D}^2 + {L \over z} F(x^i,z,u) du^2 \,.
 }
The only non-trivial component of the Einstein equations is
 \eqn{EinsteinEqs}{
  R_{uu} - {1 \over 2} g_{uu} R - {(D-1)(D-2) \over 2 L^2} g_{uu} =
    8\pi G_D J_{uu} \,,
 }
and it is straightforward to show that it takes the form
 \eqn{EinsteinSimplified}{
  \left( \square_{H_{D-2}} - {D-2 \over L^2} \right) F =
    -16 \pi G_D {z \over L} J_{uu} \,.
 }
While it may be interesting to consider the case where $F$ is non-zero over a range of $u$ (see in this connection the recent work \cite{Kajantie:2008rx}), let us restrict attention here to the case where
 \eqn{PhiAndRho}{
  F = \Phi(x^i,z) \delta(u) \qquad\hbox{and}\qquad
   J_{uu} = {L \over z} \rho(x^i,z) \delta(u) \,.
 }
Then \eno{EinsteinSimplified} becomes
 \eqn{OneSlice}{
  \left( \square_{H_{D-2}} - {D-2 \over L^2} \right) \Phi =
    -16 \pi G_D \rho \,.
 }
Choosing $\rho = E \delta(z-L) \prod_{i=1}^{D-3} \delta(x^i)$ would return us to the point-sourced shock waves that we have focused on up until now, as can be seen from comparing \eno{OneSlice} to~\eno{LaplaceShock}.  The most general $O(D-2)$-symmetric shock localized at $u=0$ corresponds to a source term $\rho$ depending only on the chordal distance variable $q$ defined in \eno{qDefAgain}: then \eno{OneSlice} becomes
 \eqn{qSlice}{
  q(1+q) \Phi'' + {1 \over 2} (1+2q)(D-2) \Phi' - (D-2) \Phi =
    -16\pi G_D L^2 \rho \,.
 }
To solve \eno{qSlice}, we follow the classic approach of first solving the homogenous equation and then using a Green's function to solve the general inhomogeneous equation.  The solutions to the homogenous equation are
 \eqn{PhiPlusMinus}{
  \Phi_-(q) &= 1 + 2q  \cr
  \Phi_+(q) &= q^{2-D} {}_2F_1(D-2,D/2;D;-1/q) \,.
 }
Note that $\Phi_-(q)$ is the unique solution that remains finite at $q=0$, and $\Phi_+(q)$ is the unique solution that decays to zero at infinity.  The Green's function $G(q,q_0)$ satisfies
 \eqn{EGreensShockSource}{
  &\left[ q(1+q) \partial_q^2 -\frac{1}{2}(1+2q)(D-2) \partial_q - (D-2)
    \right] G(q,q_0) =
   -{16\pi G_D L^2 \over [q(1+q)]^{(D-4)/2}}
    \delta(q-q_0) \,.
 }
$G(q,q_0)$ is uniquely specified by the requirement that when $q_0 >0$, $G(q,q_0)$ should be finite at $q=0$ and should decay to $0$ as $q \to \infty$.  Straightforward calculations lead to
 \eqn{Gform}{
  G(q,q_0) = {8\pi G_D L^2 \over D-1}
   \left\{ \eqalign{ \Phi_+(q_0) \Phi_-(q) &\qquad\hbox{for $q \leq q_0$}  \cr
     \Phi_-(q_0) \Phi_+(q) &\qquad\hbox{for $q \geq q_0$.}
     } \right.
 }
The solution to the original problem \eno{qSlice} is
 \eqn{GreenSoln}{
  \Phi(q) = \int_0^\infty dq_0 \,
    \left[ q_0(1+q_0) \right]^{(D-4)/2} G(q,q_0) \rho(q_0) \,.
 }
Assuming that $\rho$ has compact support, or else decays quickly enough at infinity, the asymptotic behavior of $\Phi$ near infinity is
 \eqn{PhiBigQ}{
  \Phi(q) \to {2^{6-D} \pi G_D E \over L^{D-4}(D-1)(\Vol S^{D-3})} \Phi_+(q)
    \qquad\hbox{as $q \to \infty$} \,,
 }
where we have defined
 \eqn{EdefSlice}{
  E &= 2^{D-3} L^{D-2} (\Vol S^{D-3}) \int_0^\infty dq \,
     \left[ q(1+q) \right]^{(D-4)/2} (1+2q) \rho(q)  \cr
    &= \int_{H_{D-2}} d^{D-3} x_i \, dz \left( {L \over z} \right)^{D-2}
      (1+2q) \rho(x^i,z) \,.
 }
The power of $L/z$ in the second line is from the measure on $H_{D-2}$ associated with the metric \eno{dsHD}.  The asymptotic expression \eno{PhiBigQ} for $\Phi(q)$ coincides with the solution \eno{PhiGeneralSoln} for a point-sourced shock wave.  This amounts to a sort of shell theorem: an $O(D-2)$-symmetric cloud of massless particles gives rise to the same gravitational field, outside the cloud,
as if the cloud were replaced by a single massless particle at its center with energy $E$.

\subsection{The gauge theory stress tensor of colliding shocks}
\label{MINKOWSKI}

Before proceeding to calculate the trapped surface associated with the colliding shocks, we make an aside to discuss their dual boundary theory stress energy tensor. The holographic image of a shock wave geometry on four dimensional Minkowski space has a stress-energy tensor that can be found from the small $z$ asymptotics of $\Phi$.  If  \eqn{BoundaryAsymptotics}{
  ds^2 = ds_{AdS_5}^2 + \frac{L^2}{z^2}\delta g_{mn} dx^m dx^n \,,
 }
where $\delta g_{mn} \sim {\cal O}(z^4)$ for small $z$ and has no non-zero components with a $z$ index, then
\cite{deHaro:2000xn}
 \eqn{TmnRule}{
  \langle T_{mn} \rangle = {L^3 \over 4\pi G_5} \lim_{z\to 0}
    {1 \over z^4} \delta g_{mn} \,.
 }
(This form of $\langle T_{mn} \rangle$ holds when the boundary metric is chosen to be $-dt^2 + d\vec{x}^2$.  Conformal transformations on the boundary require a modification of \eno{TmnRule}.)  We have used the notation $\delta g_{mn}$ to represent the deviation of the metric from empty \AdS even though this deviation is not necessarily small. Applying the general rule \eno{TmnRule} to the shock wave metric \eno{ShockPoincare}, one finds
 \eqn{GotTuuShock}{
  \langle T_{uu} (\vec{x}) \rangle =
   {L^2 \over 4\pi G_5} \lim_{z\to 0}
   {1\over z^3} \Phi(x^1, x^2, z) \delta(u) =
   {2 L^4 E\over \pi \left(L^2 + (x^1)^2 +
    (x^2)^2\right)^3} \delta(u)\,,
 }
with all other components vanishing when one uses the coordinate system $(u,v,x^1,x^2)$.  This same result may be obtained (as it essentially was in \cite{Horowitz:1999gf}) by first computing the stress tensor for the unboosted black hole and then applying the boost \eno{Boost} directly to $\langle T_{mn} \rangle$.

For $D \neq 5$ the analysis is similar: the energy momentum tensor associated with the metric \eqref{PoincareAnsatz} may be evaluated using
\eqref{PhiGeneralSoln} and
\begin{equation}
\label{E:TmnDRule}
	\langle T_{mn} \rangle = \frac{(D-1)L^{D+1}}{16 \pi G_D}\lim_{z \to 0}\frac{1}{z^{D-1}}\delta g_{mn} \,,
\end{equation}
the equivalent of \eqref{TmnRule} \cite{deHaro:2000xn}.  We find that
\begin{equation}
\label{E:Tuu}
	\langle T_{uu} \rangle = \frac{2^{D-2}\Gamma\left(\frac{D}{2}\right)}{\pi^{\frac{D}{2}+1}(D-2)}\frac{E L^{D-1}}{(L^2+\rho^2)^{D-2}} \delta(u)
\end{equation}
where $\rho^2 = \sum_{i=1}^{D-3} x_i^2$. Also
\begin{equation}
\label{E:TotalE}
	\int d^{D-2}x \, \langle T_{00} \rangle = E \,.
\end{equation}
The profile \eno{E:TmnDRule} respects the $O(D-2)$ symmetry discussed at the end of section~\ref{DIMENSIONS}.

For the configurations discussed in \ref{OTHER}, we learn from comparing \eno{PhiGeneralSoln} to \eno{PhiBigQ} that the dual expectation value for the gauge theory stress tensor must coincide with \eno{E:Tuu}.  This illustrates a large ambiguity in the gravity representation of some given configuration of $\langle T_{mn} \rangle$.  Such an ambiguity should not be surprising since a state in the gauge theory is by no means completely specified by the one-point function of stress tensor.  Knowledge of higher point functions of the stress tensor, and possibly of other gauge-invariant operators, would resolve such ambiguities.

Going back to $D=5$, we are eventually interested in collisions of nuclei.  So we'd like to tune $L$ and $E$ in \eqref{GotTuuShock} to resemble the energy density of a boosted nucleus as closely as possible. For a gold nucleon at rest, the energy density of the nucleus can be read off of the Woods-Saxon number density $n(x^1,x^2,x^3)$:
 \eqn{WoodsSaxonDensity}{
  \epsilon(x^1,x^2,x^3) \approx m_p n(x^1,x^2,x^3) \propto
    {1 \over 1 + e^{(|\vec{x}|-R)/a}} \,,
 }
where typical values for gold are \cite{Klein:1999qj,Adams:2004rz}
 \eqn{WoodsSaxonGold}{
  R = 6.38\,{\rm fm} \qquad a = 0.535\,{\rm fm} \,.
 }
To compare the energy density with \eqref{GotTuuShock} we need to boost \eqref{WoodsSaxonDensity}.
Consider boosting stationary, pressureless dust: before the boost,
 \eqn{Pressureless}{
  \langle T_{00}(t,x^1,x^2,x^3) \rangle = \epsilon(x^1,x^2,x^3) \,,
 }
with other components vanishing in the coordinate system $(t,x^1,x^2,x^3)$.  After the boost, the non-zero components are
 \eqn{BoostedPressureless}{
  \begin{pmatrix} \langle T_{00} \rangle & \langle T_{03} \rangle
    \\
   \langle T_{30} \rangle & \langle T_{33} \rangle
  \end{pmatrix} &=
   {1 \over 1-\beta^2} \begin{pmatrix} 1 & -1 \\ -1 & 1 \end{pmatrix}
     \epsilon(x^1,x^2,(x^3-\beta t)/\sqrt{1-\beta^2})  \cr
    &\approx
   \begin{pmatrix} 1/\beta & -1 \\ -1 & \beta \end{pmatrix}
     p_R(x^1,x^2) \delta(u)
 }
where
 \eqn{epsilonTDef}{
  p_R(x^1,x^2) \equiv
    {\beta \over \sqrt{1-\beta^2}} \int dx^3 \,
      \epsilon(x^1,x^2,x^3) \,.
 }
If the limit $\beta \to 1$ is taken with $p_R(x^1,x^2)$ held fixed (i.e.~scaling $\epsilon$ down by a factor of $\sqrt{1-\beta^2}/\beta$), then \eno{BoostedPressureless} becomes simply
 \eqn{TuuPressureless}{
  \langle T_{uu} \rangle = p_R(x^1,x^2) \delta(u) \,.
 }
Applying \eqref{BoostedPressureless} to \eqref{WoodsSaxonDensity}, we can find an expression for the $uu$ component of the stress tensor of a nucleus.

\begin{figure}
  \centerline{\includegraphics[width=7in]{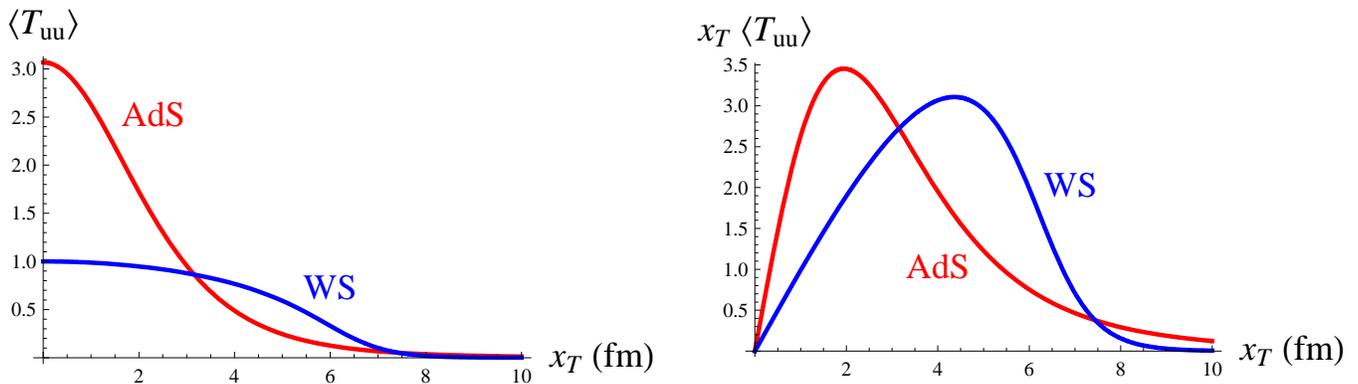}}
  \caption{Left: The dependence of $\langle T_{uu} \rangle$ on transverse radius $x_T$, both for an infinitely boosted black hole in \AdS and for an infinitely boosted Woods-Saxon profile.  $\langle T_{uu} \rangle$ is proportional to $\delta(u)$, and the quantities that we plot omit this singular factor.  The normalization of the Woods-Saxon profile was chosen so that its maximum is $1$.  The normalization and width of $\langle T_{uu} \rangle$ from the gravitational shock wave was chosen so that the integral and rms transverse radius match to the values extracted from the Woods-Saxon profile.  Right: The area under the curves $x_T \langle T_{uu} \rangle$ obtained from Woods-Saxon and AdS profiles are the same, indicating that the total energy is the same.}\label{CompareDensity}
 \end{figure}

The resulting dependence of $\langle T_{uu} \rangle$ on the transverse position $x_T = \sqrt{(x^1)^2+(x^2)^2}$ is shown in figure~\ref{CompareDensity}, together with the result \eno{GotTuuShock} obtained from the simplest gravitational shock wave.
The value of $L$ for the gravitational shock wave was chosen so that the rms transverse radius
(weighted by $p_R(x^1,x^2)$)
matches to the same quantity computed using the boosted Woods-Saxon profile.  Explicitly, for the Woods-Saxon profile,
\def\Li{\mop{Li}}
 \eqn{RMSx}{
  \langle x_T^2 \rangle &= {\int d^3 x \, [(x^1)^2+(x^2)^2]
      n(x^1,x^2,x^3) \over
    \int d^3 x \, n(x^1,x^2,x^3)}
   = {2 \over 3} {\int d^3 x |\vec{x}|^2 / (1+e^{(|\vec{x}|-R)/a})
     \over \int d^3 x / (1+e^{(|\vec{x}|-R)/a})}  \cr
   &= a \sqrt{8 \Li_5(-e^{R/a}) \over \Li_3(-e^{R/a})} \,,
 }
where $\Li_n$ are polylogarithm functions.  On the other hand, $\langle x_T^2 \rangle = L^2$ for the AdS profile $p_R(x^1,x^2) \propto [L^2+(x^1)^2+(x^2)^2]^{-3}$.  By plugging \eno{WoodsSaxonGold} into \eno{RMSx} and setting the result equal to $L^2$, one obtains $L \approx 4.3\,{\rm fm}$, as quoted in section~\ref{INTRODUCTION}.

\section{A marginally trapped surface for a head-on collision}
\label{TRAPPED}

Once the two shocks collide we can no longer superpose the solutions for two single shocks.  We assume that such a head-on collision will result in the creation of a black hole.  A standard calculation  in flat space \cite{Penrose,DEath:1992hb,DEath:1992hd,DEath:1992qu} is to estimate the area of the resulting black hole by constructing a particular trapped surface that lies on the $t<0$ parts of the $u=0$ and $v=0$ hypersurfaces.  Constructing this surface boils down to solving an unusual boundary value problem for the laplacian on the transverse space (flat ${\bf R}^2$ in the case of four-dimensional collisions).
We will follow a similar approach for shocks in anti-de Sitter space.

Heuristically, a marginally trapped surface ${\cal S}$ is the limit of a trapped surface where one of the null normals propagates inward and the other propagates in a direction that is neither inward nor outward, but only forward.  To give a more precise and useful definition, consider a null basis $(\ell^\mu,n^\mu)$ for the normal plane to ${\cal S}$ at any given point.  By convention, $\ell^\mu$ is outward pointing and $n^\mu$ is inward pointing.  Both are required to point forward in time.  ${\cal S}$ is defined by the requirement that it is closed and spacelike and that the expansion of $\ell^\mu$ should vanish:
 \eqn{thetaVanish}{
  \theta \equiv h^{\mu\nu} \nabla_\mu \ell_\nu = 0 \,,
 }
where $h_{\mu\nu}$ is the induced metric on ${\cal S}$.  The equation \eno{thetaVanish} is the mathematical expression of the heuristic notion that the outward pointing normal propagates neither outward (which would lead to positive expansion) nor inward (negative expansion).

Following \cite{Penrose,DEath:1992hb,DEath:1992hd,DEath:1992qu}, we look for a trapped surface $\cal{S}$, made up of the union of two pieces, ${\cal S} = {\cal S}_1 \cup {\cal S}_2$. The first piece ${\cal S}_1$ lies in the null hypersurface $u=0$ with $v\leq 0$ while the second piece ${\cal S}_2$ has $v=0$ and $u\leq 0$. The hypersurfaces ${\cal S}_i$ will be found by looking for appropriate codimension 2 surfaces with vanishing expansion \eqref{thetaVanish}, supplemented by the boundary condition that the outward pointing normal to $\mathcal{S}$ are continuous at the intersection $\mathcal{C} = \mathcal{S}_1\cap\mathcal{S}_2$.  A cartoon of this construction is shown in figure~\ref{3Dtrapped}, but it is important that ${\cal S}$ does not exist at a fixed time $t$: parts of it exist for all times $t<0$.\footnote{It is strange indeed to think that entropy exists at times $t<0$.  The right interpretation is that the trapped surface puts a lower bound on the amount of entropy that must eventually be created.}

Working out the shape of ${\cal S}_1$ and ${\cal S}_2$ for colliding shock waves in AdS${}_D$ is facilitated by a change of coordinates
 \eqn{CoordShift}{
  v \to v + \phi(x^i,z) \Theta(u)
 }
where $\phi(x^i,z)$ is the function appearing in \eno{PoincareAnsatz} and~\eno{PhiDef} and $\Theta(u)$ is the unit step function defined in \eqref{thetaDef}.
After this change of coordinates, the components of the metric are everywhere finite, but with jump discontinuities at $u=0$.\footnote{One can also use a more sophisticated shift, similar in spirit to the one used in \cite{Eardley:2002re} where $x_i$ and $z$ are shifted and the resultant metric becomes continuous at $u=0$. See for example \cite{Grumiller:2008va}.  This leads to the same results.}
This means that geodesics have no coordinate discontinuities.
We define ${\cal S}_1$ as the surface
 \eqn{SoneDef}{
  u = 0 \qquad v = -\psi_1(x^i,z)
 }
for some function $\psi_1$ defined on the transverse space $H_{D-2}$.  More precisely, ${\cal S}_1$ is the region of the submanifold \eno{SoneDef} where $\psi_1>0$, and its boundary ${\cal C}$ is the curve on $H_{D-2}$ (at $u=v=0$) where $\psi_1=0$.  The $O(D-2)$ symmetry of a head-on collision means that ${\cal C}$ must be a $(D-3)$-sphere, $q=q_{\cal C}$ for some constant $q_{\cal C}$.  An obvious basis for the normal space to ${\cal S}_1$ in the cotangent bundle is $(du,dv+d\psi_1)$.  So one must be able to express the outward null vector as
 \eqn{ellExpress}{
  \ell^{(1)}_\mu dx^\mu = A du + B (dv + d\psi_1) \,.
 }
Since $\ell^{(1)}_{\mu}$ is null (because $\ell^{(1)}_{\mu} {\ell^{(1)}}^{\mu}=0$), forward (because $\ell^t>0$) and outward (because $\ell^v<0$ and $v=0$ is inside the surface), we find
 \eqn{NullCombo}{
  A = -(\partial \psi_1)^2 \qquad
  B = -{4z^2 \over L^2} \,.
 }
Here $(\partial \psi_1)^2$ may be equivalently computed using the metric on $H_{D-2}$ or the metric on AdS${}_D$.

By symmetry (assuming that the momenta of the shock waves are equal and opposite), ${\cal S}_2$ must be the image of ${\cal S}_1$ under the interchange of $u$ and $v$.  Thus
 \eqn{ellExpressTwo}{
  \ell^{(2)}_\mu dx^\mu = -{1 \over 4} (\partial \psi_2)^2 dv -
    {z^2 \over L^2} (du + d\psi_2) \,,
 }
and moreover $\psi_1=\psi_2$ since the collision is head on, so let us denote them simply as $\psi$.  Continuity of the outward null normal across ${\cal C}$ means that $\ell^{(1)} = \ell^{(2)}$ when $u=v=\psi=0$.  Thus we require
 \eqn{CpsiRequire}{
  (\partial \psi)^2 = {4z^2 \over L^2}
    \qquad\hbox{on ${\cal C}$.}
 }
The Poincar\'e coordinates $(u,v,x^i,z)$ do not make the $O(D-2)$ symmetry manifest.  In place of $u$ and $v$, it is better to use $Lu/z$ and $Lv/z$, because these combinations are just $X^0-X^{D-2}$ and $X^0+X^{D-2}$, and the global coordinates $X^M$ make the $O(D-2)$ symmetry apparent.  Correspondingly we define
 \eqn{PsiDef}{
  \Psi = {L \over z} \psi \,.
 }
Evidently, \eno{CpsiRequire} together with the vanishing of $\psi$ on ${\cal C}$ are equivalent to
 \eqn{PsiBoundary}{
  \Psi|_{\mathcal{C}}=0 \qquad
    (\partial \Psi)^2|_{\mathcal{C}} &= 4 \,.
 }
Plugging either of \eno{ellExpress} or~\eno{ellExpressTwo} into \eno{thetaVanish} leads to the condition
 \eqn{PsiEquation}{
  \left( \square_{H_{D-2}} - {D-2 \over L^2} \right)
    (\Psi-\Phi) = 0 \,.
 }
The most general solution to \eno{PsiEquation} respecting the $O(D-2)$ symmetry is
 \eqn{GeneralPsi}{
  \Psi = \Phi + C\Phi_- \,,
 }
where $C$ is an integration constant, $\Phi$ is given as in \eno{PhiGeneralSoln}, and $\Phi_-(q) = 1+2q$ as in \eno{PhiPlusMinus}.  Note that $\Psi>0$ inside the curve ${\cal C}$ because $\psi$ is positive.  So \eno{PsiBoundary} becomes
 \eqn{PsiBoundarySimplest}{
  \Psi(q_{\cal C}) = 0 \qquad \Psi'(q_{\cal C}) = -{2L \over
    \sqrt{q_{\cal C}(1+q_{\cal C})}} \,.
 }
Combining \eno{GeneralPsi} with \eno{PsiBoundarySimplest} one obtains an equation for $q_{\cal C}$:
 \eqn{qCondition}{
  \Phi'(q_{\cal C}) - {2 \over 1+2q_{\cal C}} \Phi(q_{\cal C}) +
    {2L \over \sqrt{q_{\cal C}(1+q_{\cal C})}} = 0 \,.
 }
This can be conveniently rewritten as
 \eqn{WCondition}{
  W_-(q_{\cal C}) = {2 L \Phi_-(q_{\cal C}) \over
    \sqrt{q_{\cal C} (1+q_{\cal C})}}
 }
where
 \eqn{Wdef}{
  W_-(q) \equiv \Phi(q) \Phi_-'(q) - \Phi'(q) \Phi_-(q)
 }
is the Wronskian of $\Phi$ with $\Phi_-$.  Starting from \eno{qSlice}, it is easily checked that the Wronskian satisfies the following first order differential equation:
 \eqn{WdiffEQ}{
  \left[ q(1+q) \partial_q + {D-2 \over 2} (1+2q) \right] W_- =
    16\pi G_D L^2 \Phi_-(q) \rho(q) \,.
 }
The solution to \eno{WdiffEQ} is\footnote{There is a more general solution, obtained by adding a multiple of $\left[ q(1+q) \right]^{(2-D)/2}$ to \eno{WdiffSoln}.  But this solution is singular at $q=0$, and it corresponds to adding an additional, finite-energy massless particle at $q=0$.}
 \eqn{WdiffSoln}{
  W_-(q) = {\pi G_D \over 2^{D-7} L^{D-4} (\Vol S^{D-3})}
    \left[ q(1+q) \right]^{(2-D)/2} E(q) \,,
 }
where
 \eqn{EqDef}{
  E(q) \equiv 2^{D-3} L^{D-2} (\Vol S^{D-3}) \int_0^q dq_0 \,
    \left[ q_0 (1+q_0) \right]^{(D-4)/2} (1+2q_0) \rho(q_0) \,.
 }
Comparing \eno{EqDef} to \eno{EdefSlice}, one sees that $E(q)$ can be interpreted as the energy of the massless particles inside a radius $q$ in the transverse space $H_{D-2}$.  Putting \eno{WCondition} together with \eno{WdiffSoln}, we arrive at a simple relation between $q_{\cal C}$ and the energy $E_{\cal C} \equiv E(q_{\cal C})$ of massless particles inside a radius $q_{\cal C}$:
 \eqn{ECrelation}{
  {E_{\cal C} G_D \over L^{D-3}} =
   {2^{D-6} (\Vol S^{D-3}) \over \pi} (1+2q_{\cal C}) \left[ q_{\cal C}
     (1+q_{\cal C}) \right]^{(D-3)/2} \,.
 }
Table~\ref{qES}
includes values of $E_{\cal C}$ for $3 \leq D \leq 7$.
\begin{table}
 $$\seqalign{\span\TC\qquad & \span\TC\qquad & \span\TC}{
  D & E_{\cal C} G_D / L^{D-3} & S_{\rm trapped} G_D / L^{D-2}  \cr
   \hline \cr\noalign{\vskip-4\jot}
  3 & {1+2q_{\cal C} \over 4\pi} & \sinh^{-1} x_{\cal C}  \cr
  4 & \sqrt{q_{\cal C}(1+q_{\cal C})} {1+2q_{\cal C} \over 2} &
   \pi (\sqrt{1+x_{\cal C}^2} - 1)  \cr
  5 & 2 q_{\cal C} (1+q_{\cal C}) (1+2q_{\cal C}) &
   \pi (x_{\cal C} \sqrt{1+x_{\cal C}^2} -
     \sinh^{-1} x_{\cal C})  \cr
  6 & 2\pi q_{\cal C}^{3/2} (1+q_{\cal C})^{3/2} (1+2q_{\cal C}) &
   {\pi^2 \over 3} [(x_{\cal C}^2-2) \sqrt{1+x_{\cal C}^2} + 2]  \cr
  7 & {16\pi \over 3} q_{\cal C}^2 (1+q_{\cal C})^2 (1+2q_{\cal C}) &
   {\pi^2 \over 6} [ (2x_{\cal C}^2-3) x_{\cal C}
     \sqrt{1+x_{\cal C}^2} + 3 \sinh^{-1} x_{\cal C} ]
 }$$
  \caption{Values of $E_{\cal C}$ and $S_{\rm trapped}$ for $3 \leq D \leq 7$, in terms of the chordal distance variable $q_{\cal C}$ and $x_{\cal C} = 2\sqrt{q_{\cal C}(1+q_{\cal C})}$.\label{qES}}
 \end{table}

Once $q_{\cal C}$ is known, it is straightforward to calculate the area of the marginally trapped surface ${\cal S}$.
Because ${\cal S}_1$ is embedded in the null hyperplane $u=0$, whose transverse part is $H_{D-2}$ and whose lightlike direction is parameterized by $v$, the induced metric on ${\cal S}_1$ is identical to the one on $H_{D-2}$ inside a radius $q_{\cal C}$.\footnote{Explicitly: the induced metric on ${\cal S}_1$ is
 $$
  ds_{{\cal S}_1}^2 = {L^2 \over z^2} \left[ -
   {\partial u \over \partial\xi^\alpha}
    {\partial v \over \partial\xi^\beta} +
   {\partial x^1 \over \partial\xi^\alpha}
    {\partial x^1 \over \partial\xi^\beta} +
   {\partial x^2 \over \partial\xi^\alpha}
    {\partial x^2 \over \partial\xi^\beta} +
   {\partial z \over \partial\xi^\alpha}
    {\partial z \over \partial\xi^\beta} \right]
    d\xi^\alpha d\xi^\beta \,,
 $$
where $\xi^\alpha = (\xi^1,\xi^2,\xi^3)$ are any choice of coordinates on ${\cal S}_1$.  The first term drops out because $u=0$ identically.  Choosing $\xi^\alpha = (x^1,x^2,z)$, we immediately recover the metric of $H_3$.
}
Thus, the area of ${\cal S}_1$ is just the volume of the ball $q\leq q_{\cal C}$ in $H_{D-2}$, namely
 \eqn{BallArea}{
  A_{{\cal S}_1} =
   L^{D-2} (\Vol S^{D-3})
    \int_0^{x_{\cal C}} dx \frac{x^{D-3}}{\sqrt{1+x^2}} \,.
 }
In \eno{BallArea} we have introduced a new radial coordinate
 \eqn{xDef}{
  x = r/L = 2\sqrt{q(1+q)} \,,
 }
where $r$ is the radial variable appearing in \eno{GlobalToY}, and $q$ is the usual chordal distance variable, appearing for example in \eno{qDefAgain}.  The area of ${\cal S}_2$ is of course the same as of ${\cal S}_1$.  So the entropy bound is
 \eqn{HolographicEntropyBound}{
  S \geq S_{\rm trapped} \equiv
	 {1 \over 4 G_D} 2 A_{{\cal S}_1} \,.
 }
The integral \eno{BallArea} can be expressed in terms of incomplete beta functions, but the explicit form is unenlightening.  Table~\ref{qES} includes values of $S_{\rm trapped}$ for $3 \leq D \leq 7$.

In principle, $q_{\cal C}$ and $x_{\cal C}$ can be eliminated from the relations \eno{ECrelation}, \eno{BallArea}, \eno{xDef}, and \eno{HolographicEntropyBound} to obtain an explicit dependence of $S_{\rm trapped}$ on $E_{\cal C}$.  In practice, this is difficult to carry out explicitly for $D>3$.  However, when $q_{\cal C} \gg 1$, it is straightforward to extract the leading power law dependence: again for $D>3$,
\begin{equation}
\label{E:Strapped}
	S_{\rm trapped} \approx
	  C_D \left(\frac{L^{D-2}}{G_D}\right)^{1/D-2}
	  (E_{\cal C} L)^{D-3/D-2}
\end{equation}
with
 \eqn{CDDef}{
	C_D = {\left( 2^{2D-7} \pi^{D-3} \Vol S^{D-3}
	  \right)^{1 \over D-2} \over D-3} \,.
 }
When the shock wave is point-sourced, $E_{\cal C}=E$, the total energy.  The special $D=5$ case of \eno{E:Strapped}, with $E_{\cal C}=E$, was quoted in \eno{STrappedFromE}.  When calculating $S_{\rm trapped}$ in shock wave collisions intended for comparison with heavy-ion collisions, it suffices to use the leading power law dependence indicated in \eno{E:Strapped}.  The value of $S_{\rm trapped}$ obtained by using the exact parametric relations given in table~\ref{qES} are only about $0.1\%$ different when $q_{\cal C} \approx 38$, which is the value corresponding to $EL \approx 4.3 \times 10^5$.  The correction is so small because it is parametrically ${\cal O}\left( {\log q_{\cal C} \over q_{\cal C}^2} \right)$ for $D=5$.\footnote{The parametric dependence of the leading correction to \eno{E:Strapped} depends on $D$: it is ${\cal O}(1/q_{\cal C})$ for $D=4$ and ${\cal O}(1/q_{\cal C}^2)$ for $D=6$ and $D=7$.}

The case $D=3$ is evidently special.  The leading order expansions \eno{E:Strapped} and~\eno{CDDef} don't work, but instead we can eliminate $q_{\cal C}$ and $x_{\cal C}$ altogether from the expressions in table~\ref{qES} and find
 \eqn{ExplicitES}{
  S_{\rm trapped} {G_3 \over L} = \cosh^{-1} 4\pi G_3 E_{\cal C} \,,
 }
where we must have $4\pi G_3 E_{\cal C} \geq 1$ to form a trapped surface at all.  Let us focus on point-sourced shocks, so that $E_{\cal C}=E$, the total energy of one shock wave.  Remarkably, the result \eno{ExplicitES} coincides with a well-known exact result \cite{Matschull:1998rv} on light-like particles in AdS${}_3$.  The quantities denoted $p^0$ and $\epsilon$ in \cite{Matschull:1998rv} (see for example (2.9) and (3.1) of that paper) should be identified as
 \eqn{GotPzero}{
  p^0 = \tan\epsilon = 4\pi G_3 E\ \,.
 }
Using (4.3) and (4.6) of \cite{Matschull:1998rv} and noting that in this paper $L$ is set to $1$, one finds
 \eqn{FoundEntropy}{
  S = {\ell \over 4 G_3} = {L \over G_3} \cosh^{-1} p^0 \,,
 }
which indeed agrees with \eno{ExplicitES}.  Also, it was shown in \cite{Matschull:1998rv} that for $p^0 < 1$, no horizon would form in the collision of the light-like particles; instead, they would merge to form a massive particle with no horizon around it.  This matches with the observation that one needs $4\pi G_3 E \geq 1$ to form a trapped surface.

As noted in \cite{Matschull:1998rv}, black hole circumference can't increase except by sudden events like its formation, because gravity is non-dynamical in AdS${}_3$.  So it makes sense that the bound $S \geq S_{\rm trapped}$ should be saturated.  A partial translation of this statement to the dual field theory is that no entropy increase is possible after formation of a thermal state because there is no viscosity.  Bulk viscosity is forbidden by conformal invariance, and there is no shear in $1+1$ dimensions.  Similar observations have recently been made in \cite{Kajantie:2008rx,Shuryak:2008pz}.

So $S=S_{\rm trapped}$ in AdS${}_3$, at least for point-sourced shocks.  As pleasing as this result appears, we remain puzzled on a conceptual level: the field theory dual probably enjoys some form of integrability and/or holomorphic factorization which permits only forward scattering.  If one tries to collide pairs of shock waves in the dual field, forward scattering says that they should pass right through each other.  It would be nice to imagine that after the collision, the shocks ``bleed'' detritus from their leading edge, and this detritus comes to rest (on average) as a fully-formed thermal medium.  But we could not see how to describe such a process hydrodynamically without violating local momentum conservation.  It should probably be kept in mind that the exact results of \cite{Matschull:1998rv} rely on the assumption that colliding particles merge.\footnote{The additional dynamical claim of \cite{Matschull:1998rv} is that the holonomy of the end state particle is the product of the holonomies of the initial state particles.  This product rule explains how the difference between $p^0<1$ and $p^0>1$ arises: in the former case, the product of holonomies of initial state particles is a rotation, meaning that its fixed set is timelike, corresponding to a massive particle; while in the latter case, the product is a boost, corresponding to a spacelike geodesic interpreted as the future singularity inside the black hole.}  This is certainly a minimal assumption, but it does not seem inescapable.

\section{Discussion}
\label{DISCUSSION}

The scaling $S_{\rm trapped} \propto E^{2/3}$ is the most distinctive feature of point-sourced shocks in \AdS and, as we have already remarked, it will conflict with data
if the $S \propto E^{1/2}$ behavior observed in heavy ion data to date extends significantly above the scale of RHIC collisions, or if the slower increase of $S$ with $E$ predicted by CGC calculations is realized.  Let's consider how we might modify the colliding shocks so as to be consistent with a slower increase of $S$ with $E$.  Instead of asking how we could suppress entropy production while keeping energy fixed, it is intuitively easier to consider collisions with fixed $S_{\rm trapped}$ and ask how we could add energy to them.  An obvious approach is to use a halo effect: diffuse energy density can be added outside of where the trapped surface forms, and because it's so diffuse, it doesn't cause significant entropy production during the collision.  But instead of spreading out the energy density in ${\bf R}^{3,1}$, we're going to spread it out in \AdS in such a way that $\langle T_{mn} \rangle$ in the gauge theory doesn't change.  As noted in section~\ref{OTHER}, there is a large freedom in how to do this: any gravitational source in \AdS which has the $O(3)$ symmetry preserved by a massless point particle and which is localized at $u=0$ (or $v=0$) will give the same $\langle T_{mn} \rangle$ (but, presumably, with different higher point functions of $T_{mn}$ and other operators).  We will consider a particular class of $O(3)$-symmetric sources that make it easy to obtain explicit formulas using results already established.  Namely, let the gravitational stress tensor be of the form \eno{PhiAndRho} where
 \eqn{rhoCloud}{
  \rho(q) = {a E \over 8 \pi L^3 \sqrt{q(1+q)}} (1+2q)^{-2-a} \,,
 }
for some positive constant $a$. The overall normalization was chosen so that $E$ is defined as in \eno{EdefSlice}. To simplify notation, let us introduce scaled forms of the energy and entropy:
 \eqn{HattedDefs}{
  \hat{E} = {2 G_5 E \over L^2} \qquad
   \hat{S} = {G_5 S_{\rm trapped} \over \pi L^3} \,.
 }
It is helpful to keep in mind that $\hat{E}$ and $\hat{S}$ are numerically large: with $G_5$ and $L$ chosen as indicated in section~\ref{INTRODUCTION}, $E = 19.7\,{\rm TeV}$ and $S = 35000$ translates into $\hat{E} \approx 4.5 \times 10^5$ and $\hat{S} \approx 5900$.

From the third entry in table~\ref{qES} for $S_{\rm trapped}$ one can show that
 \eqn{GotShat}{
  \hat{S} \approx 4 q_{\cal C}^2 \,,
 }
where the approximate equality becomes more accurate at large $\hat{S}$.  Using \eno{EqDef}--\eno{ECrelation} together with \eno{GotShat}, one arrives at
 \eqn{GotEhat}{
  \hat{E} \approx {\hat{S}^{3/2} \over 1 - (1+\sqrt{\hat{S}})^{-a}}
    \,,
 }
where again the approximate equality becomes more accurate at large $\hat{S}$.  Taking $a \to \infty$ for fixed $\hat{S}$, one recovers from \eno{GotEhat} the leading order result $\hat{E} \approx \hat{S}^{3/2}$, which is equivalent to \eno{E:Strapped} for $D=5$.  This is because in the large $a$ limit, the distribution of energy \eno{rhoCloud} becomes pointlike in \AdS.

By allowing $a$ to be a function of $\hat{S}$, one can evidently persuade the result \eno{GotEhat} to conform to a desired scaling relation, at least for large $\hat{S}$: for example, to get $\hat{E} \approx K \hat{S}^2$ for some constant $K$, one would choose
 \eqn{aForce}{
  a = {2 / K \over \sqrt{\hat{S}} \log \hat{S}}
    \left( 1 + {1 \over 2K \sqrt{\hat{S}}} + \ldots \right)
 }
at large $\hat{S}$.  Small $a$ means that the cloud of matter in \AdS is very diffuse, so that most of the energy is in the halo where little entropy production occurs.

The strategy outlined here for obtaining a scaling $S_{\rm trapped} \propto E^{1/2}$ seems to us ad hoc.  Why would we disperse the matter in \AdS as we increase the boost factor?  Some guidance from other physical principles---perhaps saturation physics---is needed to specify more precisely what initial state we should choose in the holographic dual.
Another possibility to explain a deviation from the $E^{2/3}$ scaling is that even slight broadening of the matter distribution in the longitudinal direction would lead to a substantial reduction in the production of entropy.  This possibility is hard to assess without a more careful analysis of trapped surfaces in geometries with longitudinal smoothing.  We leave such an investigation for future work.
Of course, we should not entirely neglect the possibility that future heavy-ion data will show a markedly faster increase in total multiplicity with energy than the Landau model predicts.  If \eno{LHCmultiplicity} turns out to be about right, we would see it as evidence that the trapped surface computation captures an important aspect of the overall dynamics of the collision.

Although entropy estimates have been our main focus, the $O(3)$ symmetry that we noted in head-on collisions of point-sourced shocks has independent interest.  In this context, it is interesting to note the proposal of \cite{Iancu:2007st}, according to which the saturation scale $Q_s$
should vary across the transverse plane in a fashion similar to \eno{GotTuuShock}:
 \eqn{QsVary}{
  Q_s(x^1,x^2) = {L^2 \over L^2 + (x^1)^2 + (x^2)^2} Q_s^{\rm max} \,,
 }
where $L$ is taken to be the transverse size of the hadron under consideration.\footnote{The saturation scale is the typical transverse momentum of color field configurations in a highly boosted nucleus.  It arises when perturbative splitting causes the phase space density of gluons to become of order $1/\alpha_s$, at which point recombination of gluons cannot be neglected.  For gold-gold collisions at RHIC, an approximate value for $Q_s$ is $1\,{\rm GeV}$.  See for example \cite{McLerran:2001sr} for an introductory account of the saturation scale and related ideas.}  The form \eno{QsVary} was proposed in order to economically accommodate the known power law behavior $Q_s \sim 1/[(x^1)^2+(x^2)^2]$ at large transverse $x$ together with a finite maximum at $x^1=x^2=0$.  It is assumed to arise from feeding an initial state (the hadron at rest) which explicitly breaks conformal symmetry into evolution equations (BFKL and generalizations) which are conformally invariant, at least in the leading-log approximation.

The form \eno{QsVary} is tantalizingly similar to \eno{GotTuuShock}, hinting that the initial state and early dynamics of the collision might be closer to respecting the $O(3)$ symmetry we have found than one would expect from comparing the shapes of the Woods-Saxon and AdS profiles shown in figure~\ref{CompareDensity}. Indeed, the profile \eno{GotTuuShock} can be recognized as special even without recourse to the gauge-string duality: it is uniquely specified, up to the choice of $E$ and $L$, by invariance under the $O(3)$ remnant of the conformal group that we mentioned in section~\ref{INTRODUCTION} and identified explicitly at the end of section~\ref{BOOSTING}. We speculate that this $O(3)$ may be approximately realized in central heavy ion collisions.  If it were realized exactly in the initial state, it would be preserved during the collision to the extent that the relevant dynamics---perturbative or strongly coupled---respects conformal symmetry.

The $O(3)$ symmetry has a particularly simple realization
on the $S^3 \times {\bf R}$ boundary of global \AdS.
To make it explicit, let's introduce explicit polar coordinates in \AdS as follows:
 \eqn[c]{PolarCoords}{
  X^{-1} = \sqrt{\rho^2+L^2} \cos {\tau \over L} \qquad
  X^0 = \sqrt{\rho^2+L^2} \sin {\tau \over L}  \cr
  X^1 = \rho \sin\psi \sin\vartheta \cos\phi \qquad
  X^2 = \rho \sin\psi \sin\vartheta \sin\phi  \cr
  X^3 = \rho \cos\psi \qquad
  X^4 = \rho \sin\psi \cos\vartheta \,.
 }
The angle $\phi$ is the usual azimuthal angle around the beam-line, but the other angles do not have such a familiar interpretation: in particular, $\vartheta$ is not the angle relative to the beam.  The boundary metric is
 \eqn{BoundaryMetric}{
  ds^2 = -d\tau^2 + L^2 (d\psi^2 + \sin^2 \psi \, d\vartheta^2 +
   \sin^2 \psi \sin^2 \vartheta d\phi^2) \,.
 }
As we saw in the discussion at the end of section~\ref{BOOSTING}, it takes a massless particle a global time $\Delta\tau = \pi L$ to traverse \AdS.  Let's say that a right-moving particle starts on the boundary at $\psi=\pi$ at global time $\tau = -\pi L/2$, and a left-moving particle starts at the same time at $\psi=0$.  The propagation of these particles toward one another is dual to an expansion of light-like, spherically symmetric shock waves from the insertion points at $\psi=\pi$ and $0$.  At times $-\pi L/2 < \tau < 0$, the stress tensor on $S^3 \times {\bf R}$ is
 \eqn{EarlyStress}{
  \langle \tilde{T}_{\tau\tau} \rangle &=
   {1 \over L^2} \langle \tilde{T}_{\psi\psi} \rangle =
    {E \over 4\pi L^3 \sin^2 \psi} \left[
     \delta\left( \psi + {\tau \over L} - {\pi \over 2} \right) +
     \delta\left( \psi - {\tau \over L} - {\pi \over 2} \right)
    \right]  \cr
  \langle \tilde{T}_{\tau\psi} \rangle &=
   \langle \tilde{T}_{\psi\tau} \rangle =
    {E \over 4\pi L^2 \sin^2 \psi} \left[
     -\delta\left( \psi + {\tau \over L} - {\pi \over 2} \right) +
     \delta\left( \psi - {\tau \over L} - {\pi \over 2} \right)
    \right] \,,
 }
with other components vanishing.
The first term in square brackets of each of the explicit expressions in \eno{EarlyStress} is due to the right-moving shock, and the second term to the left-moving shock.  The notation $\tilde{T}_{ab}$ for the stress tensor on $S^3 \times {\bf R}$ reminds us not only that there is a non-trivial conformal mapping between this form and \eno{GotTuuShock}, but also that the stress tensor picks up an anomalous vacuum contribution, proportional to the metric, in the course of this mapping, which we have excluded from $\tilde{T}_{ab}$.

It would be interesting to start from \eno{EarlyStress}, or some alteration of it where the delta functions are softened into sharply peaked but smooth functions, and evolve it forward with hydrodynamics, or with some combination of a heuristic treatment of thermalization for $0 < \tau < \tau_{\rm therm}$ followed by hydrodynamics.  Although such evolution would be a purely one-dimensional problem, it would incorporate a combination of radial and longitudinal flow in the original Minkowski-space conformal frame.

It is worth noting that an $O(2)$ remnant of the conformal group is preserved even when the collision is not head-on. (If $D > 5$, then this would be an $O(D-3)$ symmetry.) This is easiest for us to see by considering test particles in \AdS.  In Poincar\'e coordinates, the right-moving particle travels on a trajectory with
 \eqn{PoincareRight}{
  x^3 = t \qquad x^1=x^2=0 \qquad z=L \,,
 }
and the left-moving particle's trajectory is
 \eqn{PoincareLeft}{
  x^3 = -t \qquad x^1=0 \qquad x^2 = b \qquad z=L \,,
 }
where $b$ is the impact parameter.  In global coordinates, these trajectories take the form
 \eqn{GlobalLR}{\seqalign{\span\TT &\qquad \span\TR}{
  right-moving: & X^3=X^0 \qquad X^1=X^2=X^4=0 \qquad
    X^{-1} = L  \cr
  left-moving: & \left\{ \eqalign{ X^3&=-X^0 \qquad X^1=0
     \qquad X^2=b  \cr\noalign{\vskip-2\jot}
    X^4 &= -{b^2 \over 2L} \qquad X^{-1} = L + {b^2 \over 2L} \,.}
   \right.
 }}
The total configuration \eno{GlobalLR} is invariant under the full $O(3)$ subgroup iff $b=0$, but for $b \neq 0$ it is still invariant under the $O(2)$ subgroup that preserves the vector $(X^1,X^2,X^4) = (0,b,-b^2/2L)$.  Because the dynamics respects the full conformal symmetry, the final state should be invariant under $O(3)$ or $O(2)$, according as $b=0$ or $b \neq 0$.

Clearly, estimating total entropy production is only one facet of describing colliding shocks in anti-de Sitter space.  Ideally, one would like to understand the process of thermalization and the subsequent hydrodynamical flow.  In the case of central collisions, imposing the $O(3)$ symmetry means that the gravity calculations are effectively $2+1$-dimensional.  An optimistic view is that a fairly full account could be made by somehow matching a marginally trapped
surface computation to a late-time description that fuses a linearized treatment of non-hydrodynamical quasi-normal modes, as described in \cite{Kovtun:2005ev,Friess:2006kw}, with a non-linear treatment of hydrodynamical modes, as described in \cite {Bhattacharyya:2008jc}.  In such an account, the local thermalization time might be expected to be several times the relaxation time of the non-hydrodynamical modes: in total, roughly $0.3\,{\rm fm}/c$ for
conditions comparable to central RHIC collisions, according to the estimate of \cite{Friess:2006kw}.  An alternative approach to describing thermalization following a collision of two shocks in \AdS has been suggested in \cite{Grumiller:2008va}.

\section*{Acknowledgements}

We thank S.~Klainerman, M.~Lublinsky, E.~Witten, and W.~Zajc for discussions and suggestions.  A.Y.~thanks the Institute for Nuclear Theory at the University of Washington for its hospitality and the Department of Energy for partial support during the completion of this work. The work of S.S.G.~and S.S.P.~was supported in part by the Department of Energy under Grant No.\ DE-FG02-91ER40671 and by the NSF under award number PHY-0652782.  A.Y. is supported in part by the German science foundation and by the Minerva foundation.

\bibliographystyle{ssg}
\bibliography{shocks}

\end{document}